\documentclass[aps,pra,reprint,superscriptaddress,showpacs]{revtex4-1}

\usepackage{amsmath}
\usepackage{amsfonts}
\usepackage{amssymb}
\usepackage{graphicx}
\usepackage{graphics}
\usepackage{hyperref}

\begin{document}


\title{Finite geometry models of electric field noise from patch potentials in ion traps}


\author{Guang Hao Low}
\affiliation{MIT-Harvard Center for Ultracold Atoms, Department of Physics,
Massachusetts Institute of Technology, Cambridge, Massachusetts 02139, USA}
\affiliation{Department of Physics, Cavendish Laboratory, J. J. Thomson Avenue, Cambridge, CB3 0HE, UK}
\author{Peter F. Herskind}
\affiliation{MIT-Harvard Center for Ultracold Atoms, Department of Physics,
Massachusetts Institute of Technology, Cambridge, Massachusetts 02139, USA}
\author{Isaac L. Chuang}
\affiliation{MIT-Harvard Center for Ultracold Atoms, Department of Physics,
Massachusetts Institute of Technology, Cambridge, Massachusetts 02139, USA}


\date{\today}

\begin{abstract}
We model electric field noise from fluctuating patch potentials on conducting surfaces by taking into account the finite geometry of the ion trap electrodes to gain insight into the origin of anomalous heating in ion traps. The scaling of anomalous heating rates with surface distance, $d$, is obtained for several generic geometries of relevance to current ion trap designs, ranging from planar to spheroidal electrodes. The influence of patch size is studied both by solving Laplace's equation in terms of the appropriate Green's function as well as through an eigenfunction expansion. Scaling with surface distance is found to be highly dependent on the choice of geometry and the relative scale between the spatial extent of the electrode, the ion-electrode distance, and the patch size. Our model generally supports the $d^{-4}$ dependence currently found by most experiments and models, but also predicts geometry-driven deviations from this trend.
\end{abstract}


\maketitle



\section{Introduction \label{Introduction}}
Metallic conductors are widely considered to be ideal electric equipotentials.  However, from the perspective of precision measurements and quantum information science, real metals are surprisingly poor equipotentials.  Not only do metallic surfaces exhibit static, inhomogeneous, microscopic potential patches ~\cite{Camp1991,Camp1992,Rossi1992}, but even more interestingly, the electric fields produced by real conductors exhibit time-varying fluctuations.  These effects have profound implications for many modern experiments: the static patches produce static electric fields, which influence precision measurements ~\cite{Sukenik1993,Obrecht2007,Speake2003}, while the time-varying fluctuations, or electric field noise, limit progress on single spin detection~\cite{Mamin2003}, nanomechanics~\cite{Stipe2001,Kuehn2006,Li2007}, detection of weak forces~\cite{Lockhart1977,Speake2003,Robertson2006}, and quantum information processing with trapped ions~\cite{Turchette2000,Deslauriers2006,Epstein2007,Labaziewicz2008}. As such, many fields of inquiry would be advanced if the physical origin of these phenomena were to be understood.

A few physical models for patch potentials and electric field noise exist. The static component of patch potentials has been studied in great detail and is believed to be related to work function differences between crystal facets and to surface adsorbates~\cite{Camp1991,Rossi1992}. Electric field noise, however, is less understood, with two main candidate models. One model explains that this noise is a consequence of the finite electrical resistance of the metallic electrodes and associated external circuitry used in experiments, otherwise known as Johnson noise~\cite{Henkel1999,Henkel1999a,Carminati1999,Turchette2000,Leibrandt2007}.  A second model attributes the noise to patch potentials with a fluctuating component ~\cite{Turchette2000,Dubessy2009}, and is motivated by experimental evidence which substantiates a $1/\omega$ frequency scaling~\cite{Turchette2000,Deslauriers2006,Labaziewicz2008}, with an origin rooted in thermally activated processes~\cite{Stipe2001,Deslauriers2006,Labaziewicz2008,Labaziewicz2008a}. The fluctuating patch potential noise model is the focus of this paper.

Differentiating between noise models requires accurate and precise measurements of noise. Depending on the probing frequency and the distance from the surface at which the electric field noise is measured, very different methods may be used, such as cantilevers ~\cite{Stipe2001,Kuehn2006,Li2007}, Kelvin probes ~\cite{Camp1991,Robertson2006}, or trapped ions~\cite{Turchette2000,Deslauriers2006,Epstein2007,Labaziewicz2008}. In particular, atomic ions, cooled to the motional ground state of their trapping potential, exhibit remarkable sensitivity to electric field noise. Experiments can discern changes in the motion of the ions to within a fraction of single quanta due to extremely narrow electronic transitions in ions, which can feature fractional linewidths of about one part in $10^{14}$. This sensitivity has revealed an anomaly in the electric field produced by conductors: it is significantly noisier than predicted by Johnson noise, leading to a phenomenon known as \emph{anomalous} ion heating~\cite{Turchette2000,Deslauriers2006,Epstein2007,Labaziewicz2008,Labaziewicz2008a}.

\begin{figure}
\includegraphics[width=0.8\columnwidth]{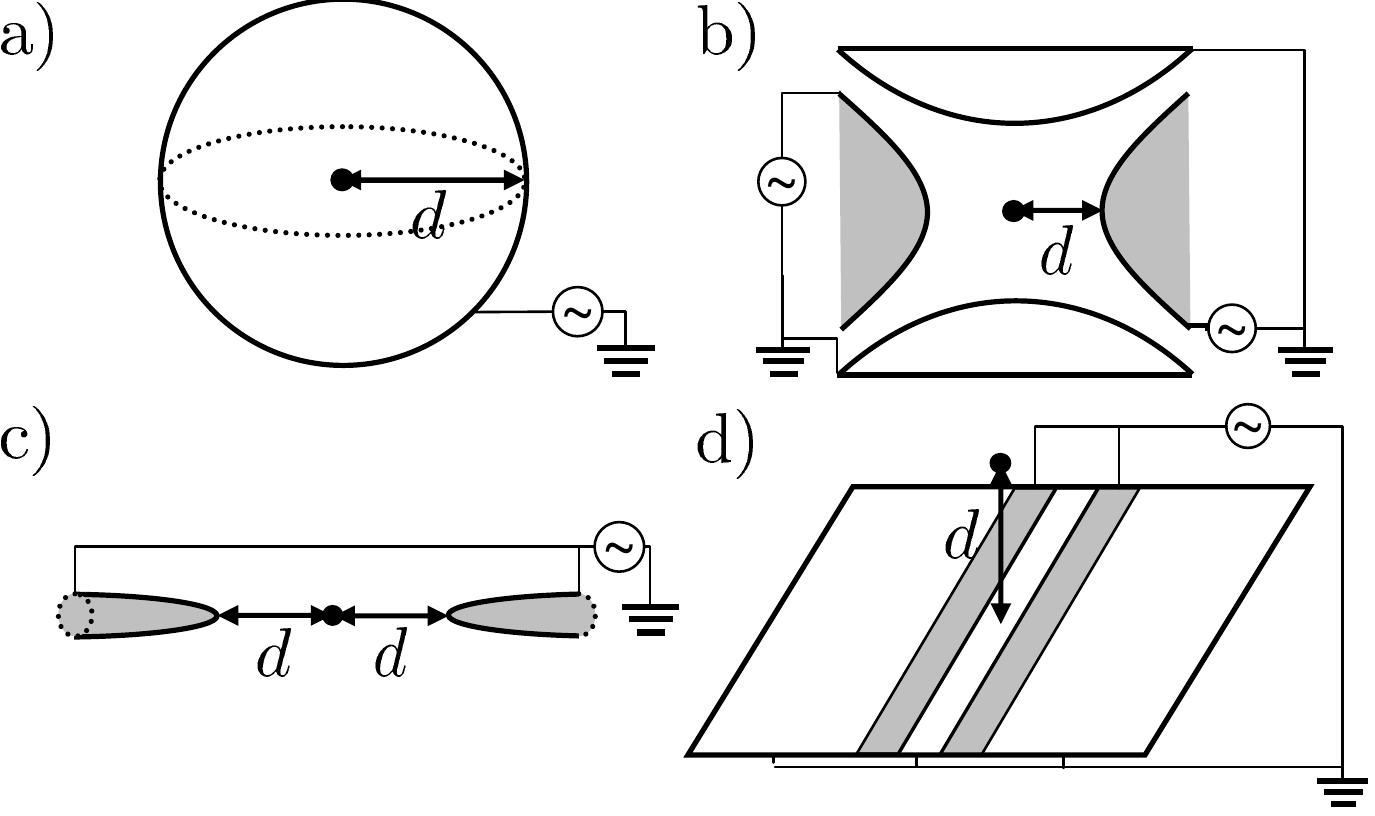}
\caption{Examples of electrode geometries relevant to ion trapping. $d$ is the nearest ion-electrode surface distance. Electrodes are at either RF potential (shaded) or DC potential (unshaded). a) Ion-in-a-sphere model used in Ref.~\cite{Turchette2000}. b) Cross sectional view of the hyperbolic Paul trap. c) Needle electrode Paul trap.  d) Surface electrode Paul trap.
\label{fig:planar_trap}}
\end{figure}

Anomalous heating in ion traps has been studied, e.g., by Turchette et al.~\cite{Turchette2000} by modeling the ion as a single charge at the center of a conducting sphere [Fig.~\ref{fig:planar_trap}~(a)], which is a reasonable approximation to, e.g., the hyperbolic radio frequency (RF) Paul trap [Fig.~\ref{fig:planar_trap}~(b)]. Their work showed that the power of electric field noise should scale with ion-electrode (conductor) distance $d$ as $d^{-\alpha}$ with a scaling exponent $\alpha=2$  for Johnson noise, and $\alpha=4$ for patch potential noise. From collective data gathered by several independent ion trap experiments~\cite{Epstein2007}, a pattern in support of the patch potential model has emerged; however, comparison between experiments is fraught with uncertainty as anomalous heating is known to be strongly influenced by trap preparation ~\cite{Turchette2000,Labaziewicz2008a}. A systematic study of the distance scaling in a single ion trap was performed by Deslauriers et al.~\cite{Deslauriers2006}, who found $\alpha=3.5\pm0.1$. While in reasonable agreement with the model of fluctuating potential patches, the needle-shaped electrode geometry of their ion trap [Fig.~\ref{fig:planar_trap}~(c)] bore little resemblance with that of the ion-in-a-sphere model of Turchette et al.

It is thus of interest to model the scaling of anomalous heating in the patch potential model for different geometries. In this respect, the Turchette et al. model is complimented by the model of Dubessy et al.~\cite{Dubessy2009}, who considered an infinite planar surface. The infinite planar geometry is representative of the technologically important surface electrode ion trap~\cite{Chiaverini2005a, Wesenberg2008, House2008}, where the ion is trapped above the surface of two-dimensional electrodes [Fig.~\ref{fig:planar_trap}~(d)], as well as of the situation encountered in non-contact friction measurements with cantilevers~\cite{Stipe2001}. Dubessy et al.~\cite{Dubessy2009} recognized the importance of modeling finite correlations between potential patches and introduced a geometric factor, $\zeta$, for the patch correlation length, which led to a prediction of a strong dependence of $\alpha$ on this length scale. Specifically, they obtained $\alpha=4$ for $d\gg\zeta$ and $\alpha=1$ for $d\ll\zeta$, which allowed for an impressive connection between data from ion trap to cantilever experiments across more than three orders of magnitude in scale.

Here, we study the influence of electrode geometry on patch potential driven electric field noise in ion traps by extending previous theoretical work~\cite{Turchette2000,Dubessy2009,Safavi-Naini2011}, to allow for a consideration of an arbitrary geometry and patch correlation function. By applying the method of eigenfunction expansions, we are led to an intuitive understanding of the effects of arbitrary patch distributions and we apply our model to explicit examples of simple finite geometries where the scaling exponent $\alpha$ is evaluated. In these simple finite geometries, we emphasize the limiting cases of infinite and infinitesimal patch sizes. This enables us to establish bounds on $\alpha$ across a range of relative scales of patch size, ion-electrode distance, and electrode dimensions.

The specific electrode geometries we consider include the infinite plane as well as finite spheroidal electrodes. The planar geometry is representative of typical surface electrode ion traps~\cite{Chiaverini2005a,Wesenberg2008,House2008} and is presented, in part, to connect with previous works~\cite{Dubessy2009,Safavi-Naini2011}. Spheroids, on the other hand, are examples of finite geometries, where limiting cases can be chosen to approximate both finite planar electrodes and the needle trap of Deslauriers et al.~\cite{Deslauriers2006}.

We find that $\alpha$ assumes a wide range of values depending on the relative scales of patch size, ion-electrode distance, and spatial extent of the electrode. Near electrode surfaces, the scaling exponent is bounded within the range $0<\alpha<4$, depending on patch size. In typical ion trap configurations with a large spatial extent of electrodes relative to an ion height of $d\sim 100~\mu$m and with small patch sizes of dimension $\sim 1~\mu$m, $\alpha$ approaches $4$, in agreement with the models of Turchette et al. and Dubessy et al. As the distance, $d$, is decreased we find that $\alpha$ decreases, as was also determined by Dubessy et al. For the special case of the planar hole trap~\cite{Brewer1992,DeVoe2002}, with the ion suspended at the center of a hole in the RF electrode, we find that the bounds are narrowed to $2< \alpha < 4$ but the $\alpha=4$ scaling is retained in the limit of typical trap dimensions. In the case of spheroidal shapes, $\alpha$ converges to either $4$ or $6$, when the ion is far from the spheroid, depending on whether the mode of motion considered is normal or transverse to its surface. For the limiting case of a needle in the intermediate regime of $d\sim a$, geometry and patch size dependence may allow for scaling consistent with the $\alpha=3.5$ value for the normal mode of motion, as found by the Deslauriers et al. experiment~\cite{Deslauriers2006}.

Our formalism and its results are presented in the following. We begin, in section~\ref{sec:Model} by briefly reviewing the general theory for heating rates in ion traps due to fluctuating patch potentials in the framework of Laplace's equation and its Green's function solution. In section~\ref{sec:GeoDep} we study the influence of electrode geometry on the scaling of heating rates for various generic geometries of relevance to current ion trap designs. Finally, in section~\ref{sec:Conclusions} we summarize key results of our work and propose directions for further studies, where the analytical methods developed here may aid in numerical simulations of non-generic geometries to model more realistic experimental scenarios.


\section{Model} \label{sec:Model}
We analyze the scaling laws behind heating rates as follows: In section \ref{sec:Model_heatrate} the dependence of single ion heating rate on fluctuating electrical fields is determined. In section \ref{sec:Model_EFN} it is assumed that these fluctuating electric fields originate from potentials on some conducting surface with a certain geometry. In section \ref{sec:Model_patchpotapprox} the patch potential approximation is implemented, which allows all information about electrode and patch geometry to be described by a single geometric factor, which we shall denote $\Lambda$. In section \ref{sec:Model_patchsize} assumptions about patch sizes are introduced to evaluate upper and lower bounds for the scaling laws. Finally, in section \ref{sec:Model_scaling} the scaling exponent $\alpha$ is derived from the geometric factor.

\subsection{Ion heating rate} \label{sec:Model_heatrate}
Our treatment of electric field noise in ion traps assumes that the ion is confined in a harmonic potential in three dimensions, where the strength of the potential is quantified by  frequencies $\omega_k$, with $k$ denoting the principle axes of the potential. Such confinement may by achieved either by a combination of static electric and magnetic fields, as in the case of the Penning trap, or by a combination of static and time-varying electric fields, as in the case of the Paul trap~\cite{Ghosh1995}. Our focus is on the Paul trap, of which a few examples are shown in Fig.~\ref{fig:planar_trap} b)--d), but our formalism is quite general, and applies to any scenario where a particle of mass $m$ and charge $q$ is held in a harmonic potential near a conducting surface. For this reason, we shall not review the subject of ion confinement here; we refer the reader to Refs.~\cite{Wesenberg2008,House2008,Ghosh1995} for detailed treatment of that subject.

We assume the ion is initially is the quantum mechanical ground state of the harmonic oscillator potential and consider the effect of a fluctuating electric field with components $E_{k}(t)$ at the ion location. The heating rate is now defined as the rate at which this field induces transitions from the ground state to the first excited state and can be evaluated via first order perturbation theory to~\cite{Turchette2000}
\begin{equation}
\Gamma_{0\rightarrow 1} =\frac{q^2}{4m\hbar\omega_k}S_{E_k}(\omega_k).
\end{equation}
Here
\begin{equation}\label{eq:SEomega}
S_{E_k}(\omega_k)\equiv 2\int^{+\infty}_{-\infty} d\tau e^{i \omega_k \tau}\langle E_k(t)E_k(t+\tau) \rangle
\end{equation}
is the power spectrum of the electric field noise and $\langle \cdots \rangle$ denotes time averaging. Cross coupling between the noise and the RF drive field at frequency $\Omega$ occurs in Paul traps, but has negligible effect as long as $(\omega_{k}/\Omega)^{2} \ll 1$~\cite{Turchette2000} and is hence omitted from our treatment.

\subsection{Electric field noise spectral density} \label{sec:Model_EFN}
\begin{figure}
\includegraphics[width=0.8\columnwidth]{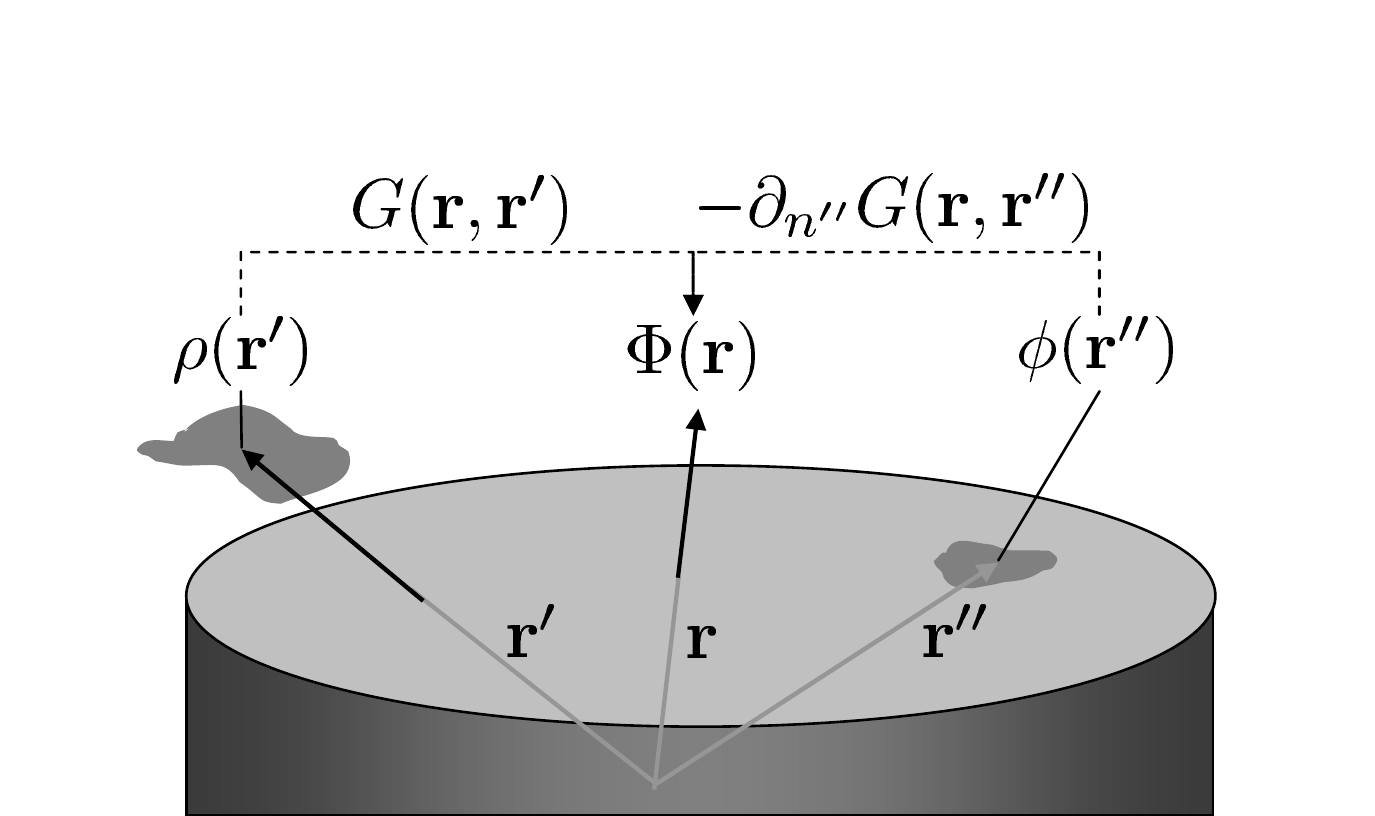}
\caption{Graphic illustration of Eq.~\ref{eq:potential} for a given, arbitrary conducting surface $\sigma$, here represented by the shaded cylinder. The electrostatic potential $\Phi(\mathbf{r})$ at point $\mathbf{r}$ is determined by adding the convolution of some charge density $\rho(\mathbf{r}')$ with the Green's function $G(\mathbf{r},\mathbf{r}')$ and the convolution of the surface potential $\phi(\mathbf{r}'')$ with the surface Green's function $-\partial_{n''}G(\mathbf{r},\mathbf{r}'')$. \label{fig:greens_function}}
\end{figure}
Literature on patch potentials~\cite{Camp1991,Camp1992,Rossi1992} suggests various origins such as crystal grain boundaries, which give rise to variations in local work functions on the conductor, and adsorbed elements on the surface, which may alter the surface potential locally as well. Sub-surface defects have also been considered; however, measurements with ions trapped above superconducting surfaces that can provide shielding against fields from such defects, suggests that their influence is weak~\cite{Wang2010b}. In the following, we thus focus on a description based on a surface effect.

The fluctuating electric field $E_{k}(t)$ (Eq.~\ref{eq:SEomega}) can be obtained by solving Laplace's equation with Dirichlet boundary conditions specified by the potential $\phi$, which is determined by fluctuating patch potentials on a surface $\sigma$ of interest. In the Green's function formalism for a time-dependent electrostatic charge density, $\rho(\mathbf{r},t)$, the potential is given by
\begin{equation}\label{eq:potential}
\Phi(\mathbf{r},t) = \int_{V} G(\mathbf{r},\mathbf{r}') \rho(\mathbf{r}',t) d\mathbf{r}'- \int_{\sigma}\phi(\mathbf{r''},t)\frac{\partial  G(\mathbf{r},\mathbf{r}'')}{\partial n''} d\mathbf{r}'',
\end{equation}
where $ G(\mathbf{r},\mathbf{r}')$ is the Green's function of Laplace's equation for some arbitrary conducting surface. $V$ is the volume of integration, and $n''$ is normal to the surface of integration $\sigma$ at some point $\mathbf{r}'' \in \sigma$ (Fig.~\ref{fig:greens_function}). We have applied here the quasi-static approximation as typical ion-electrode distances of $d\sim100~\mu$m are much less than the relevant wavelength of electric fields at the secular motion frequencies, which are typically of order $\sim10$~MHz.

As we are most interested in surface effects, $\rho$ represents charges adsorbed onto the surface, whereas $\phi$ represents patch potentials. However, the electric potential of a charge very close to a conducting surface is indistinguishable from an appropriately chosen patch potential as long as the ion-surface distance is much larger than the adsorbed charge-surface distance. Hence, for simplicity, we set $\rho=0$ and treat all sources in terms of effective patch potentials. The equivalence between surface and free-space sources implies that, for sources $\phi$ on the surface, $-\partial_{n'} G(\mathbf{r},\mathbf{r}')|_{\mathbf{r}'\in\sigma}$ is analogous to $G(\mathbf{r},\mathbf{r}')$ for sources $\rho$ in free space. We therefore define the surface Green's function $G_{\sigma}(\mathbf{r},\mathbf{r}')\equiv-\partial_{n'}G(\mathbf{r},\mathbf{r}')|_{\mathbf{r}'\in\sigma}$. In this notation, the temporal correlation function of the electric field along the $k^\mathrm{th}$ mode is
\begin{eqnarray}\label{eq:EtimeCorrelation}
\langle E_k(t)E_k(t+\tau) \rangle &=& \langle \nabla_k \Phi(\mathbf{r},t) \nabla_k \Phi(\mathbf{r},t+\tau) \rangle \nonumber\\
&=&\int_{\sigma''} \int_{\sigma'} \langle\phi(\mathbf{r}',t)\phi(\mathbf{r}'',t+\tau)\rangle \nonumber \\
&\times& \Big(\nabla_k G_{\sigma}(\mathbf{r},\mathbf{r}')\cdot \nabla_k G_{\sigma}(\mathbf{r},\mathbf{r}'')\Big) d\mathbf{r}' d\mathbf{r}'',\nonumber \\
\end{eqnarray}
where $\nabla_{k}$ is the $k^\mathrm{th}$ component of the gradient operator. Inserting into Eq.~\ref{eq:SEomega} we find
\begin{eqnarray}\label{eq:SEomega2}
S_{E_k}(\omega_k) &=& 2\int_{\sigma''} \int_{\sigma'} \mathbf{F}\left[\langle\phi(\mathbf{r}',t)\phi(\mathbf{r}'',t+\tau)\rangle\right] \nonumber \\
&\times& \Big(\nabla_k G_{\sigma}(\mathbf{r},\mathbf{r}')\nabla_k G_{\sigma}(\mathbf{r},\mathbf{r}'')\Big) d\mathbf{r}' d\mathbf{r}'',
\end{eqnarray}
where $\mathbf{F}\left[\cdots\right]=\int_{-\infty} ^\infty \left[\cdots\right]e^{i\omega \tau}d\tau$ denotes the temporal Fourier transform.

\subsection{Patch potential approximations} \label{sec:Model_patchpotapprox}
In order to evaluate Eq.~\ref{eq:SEomega2}, some assumptions need to be made about the nature of the sources. One common assumption, such as in Ref.~\cite{Dubessy2009}, is that the temporal and spatial variation of the sources decouple. In this approximation, the sources are described by a superposition of $N$ separate patches, where the $i^\mathrm{th}$ patch is described by a time dependent function $V_{i}(t)$ and an effective spatial extent $\chi_i(\mathbf{r})$ -- hence the term patch potential. Thus,
\begin{equation}
\phi(\mathbf{r},t) = \sum^N_{i=1} V_{i}(t) \chi_{i}(\mathbf{r}),
\end{equation}
where the normalization of $\chi_{r}(\mathbf{r})$ is with respect to the area $A$ of the surface $\sigma$:
\begin{equation}
\int_{\sigma} \sum^{N}_{i=1}\chi_{i}(\mathbf{r}) d\mathbf{r} = A.
\end{equation}
A measure of the average patch size may be obtained through the spatial correlation function
\begin{equation}
C(\mathbf{r}',\mathbf{r}'')=\sum^{N}_{i=1}\chi_{i}(\mathbf{r}')\cdot\chi_{i}(\mathbf{r}''),
\end{equation}
with normalization given by
\begin{equation}\label{eq:CorrelationNormalization}
\int_{\sigma'}\int_{\sigma} C(\mathbf{r}, \mathbf{r}') d\mathbf{r} d\mathbf{r}' = \frac{A^{2}}{N}.
\end{equation}
This equality is satisfied for both patches that overlap maximally and disjoint patches with zero overlap. The former case is unphysical for naturally occurring patches, but could be reached experimentally by applying strong electric noise from a single source across all surfaces. Lastly, the sources are assumed to share the same spectral distribution, $R(\omega)$, but are also uncorrelated with each other, such that
\begin{equation}
\mathbf{F}\left[\langle V_{i}(t)\cdot V_{j}(t+\tau)\rangle\right]=R(\omega)\delta_{ij}.
\end{equation}
The noise spectrum $R(\omega)$ has been observed by several experiments to resemble a $1/\omega$ scaling~\cite{Turchette2000,Deslauriers2006,Epstein2007,Labaziewicz2008}, although theoretical models exists which supports a frequency dependence extending to both $1/\omega^{1.5}$~\cite{Gesley1985} and $1/\omega^{2}$ ~\cite{Safavi-Naini2011}, depending on the assumptions used.

Given our assumptions above, the frequency dependence in the expression for the spectral density of Eq.~\ref{eq:SEomega2} is conveniently separated from the spatial dependance and becomes
\begin{eqnarray}\label{eq:SEomega3}
S_{E_k}(\omega_k) &=& 2R(\omega_{k})\int_{\sigma''} \int_{\sigma'} \left[\sum^{N}_{i=1}\chi_{i}(\mathbf{r}')\cdot \chi_{i}(\mathbf{r}'')\right] \nonumber \\
&\times& \Big(\nabla_k G_{\sigma}(\mathbf{r},\mathbf{r}')\nabla_k G_{\sigma}(\mathbf{r},\mathbf{r}'')\Big) d\mathbf{r}' d\mathbf{r}''\nonumber\\
&=& 2R(\omega_{k})\Lambda_{k}(\mathbf{r}),
\end{eqnarray}
where we have defined the geometric factor
\begin{eqnarray}\label{eq:geometricfactor}
\Lambda_{k}(\mathbf{r}) &=& \int_{\sigma''} \int_{\sigma'} C(\mathbf{r}', \mathbf{r}'')\nonumber \\
&\times& \Big(\nabla_k G_{\sigma}(\mathbf{r},\mathbf{r}')\nabla_k G_{\sigma}(\mathbf{r},\mathbf{r}'')\Big) d\mathbf{r}' d\mathbf{r}'',
\end{eqnarray}
which encompasses information about patch sizes through $C(\mathbf{r}', \mathbf{r}'')$ and the surface geometry through the surface Green's function $G_{\sigma}(\mathbf{r},\mathbf{r}')$.

\subsection{Patch size} \label{sec:Model_patchsize}
It is evident from Eq.~\ref{eq:geometricfactor} that $G_{\sigma}(\mathbf{r},\mathbf{r}')$ and $C(\mathbf{r}', \mathbf{r}'')$ are generally not separable in the expression for the power spectrum of the electric field noise. Furthermore, analytic expressions for the geometric factor do not exist for most choices of geometries or correlation functions. We deal with this issue in two ways: One, by considering limiting cases of the correlation function, and two, by considering an eigenfunction expansion of the correlation function.

The first approach takes two limiting cases of patch size relative to the spatial extent of the surface where the problem simplifies, which provides some useful intuition about the behavior encountered in intermediate regimes. The limiting cases are the Infinite Patch (IP), which corresponds to the trivial case of a conductor held at uniform potential, and the Point Patch (PP), represented by an infinitesimally small patch disjoint from all other patches. Formally,
\begin{equation}\label{eq:CorrelationFunctionIPPP}
C(\mathbf{r}', \mathbf{r}'')=\begin{cases}1 &\text{IP,} \\ \frac{A}{N} \delta^{2}(\mathbf{r}'-\mathbf{r}'')|_{\sigma} &\text{PP.}\end{cases}
\end{equation}
The choice of $A/N$ as the coefficient for the PP limit satisfies the normalization in Eq.~\ref{eq:CorrelationNormalization}, and $N$ is formally given by $A\cdot\lim_{\mathbf{r}\rightarrow 0} \delta^{2}(\mathbf{r})$. In these limits, the geometric factor takes the following forms:
\begin{equation}\label{eq:SEomega4}
\Lambda_{k}(\mathbf{r}) = \begin{cases}
\left|\int_{\sigma'}\nabla_k G_{\sigma}(\mathbf{r}, \mathbf{r}') d\mathbf{r}'\right|^{2}  &\text{IP,} \\
 \frac{A}{N}\int_{\sigma'} \left|\nabla_k G_{\sigma}(\mathbf{r}, \mathbf{r}')\right|^{2} d\mathbf{r}' & \text{PP.}\end{cases}
\end{equation}
The physical intuition behind Eq.~\ref{eq:SEomega4} is that within a large (infinite) patch, noise from all sub-elements of the patch are correlated and the fields add coherently, while for small (point) patches, the contributions from individual points are uncorrelated and the fields are added incoherently by summing the intensities. As long as patch sizes are considerably smaller than the variation of the surface geometry, this leads us to expect a smaller magnitude of noise for smaller patches than for very large patches, in agreement with the conclusions of previous models of noise originating from patch potentials~\cite{Turchette2000,Dubessy2009}.

\begin{figure}
\includegraphics[width=0.8\columnwidth]{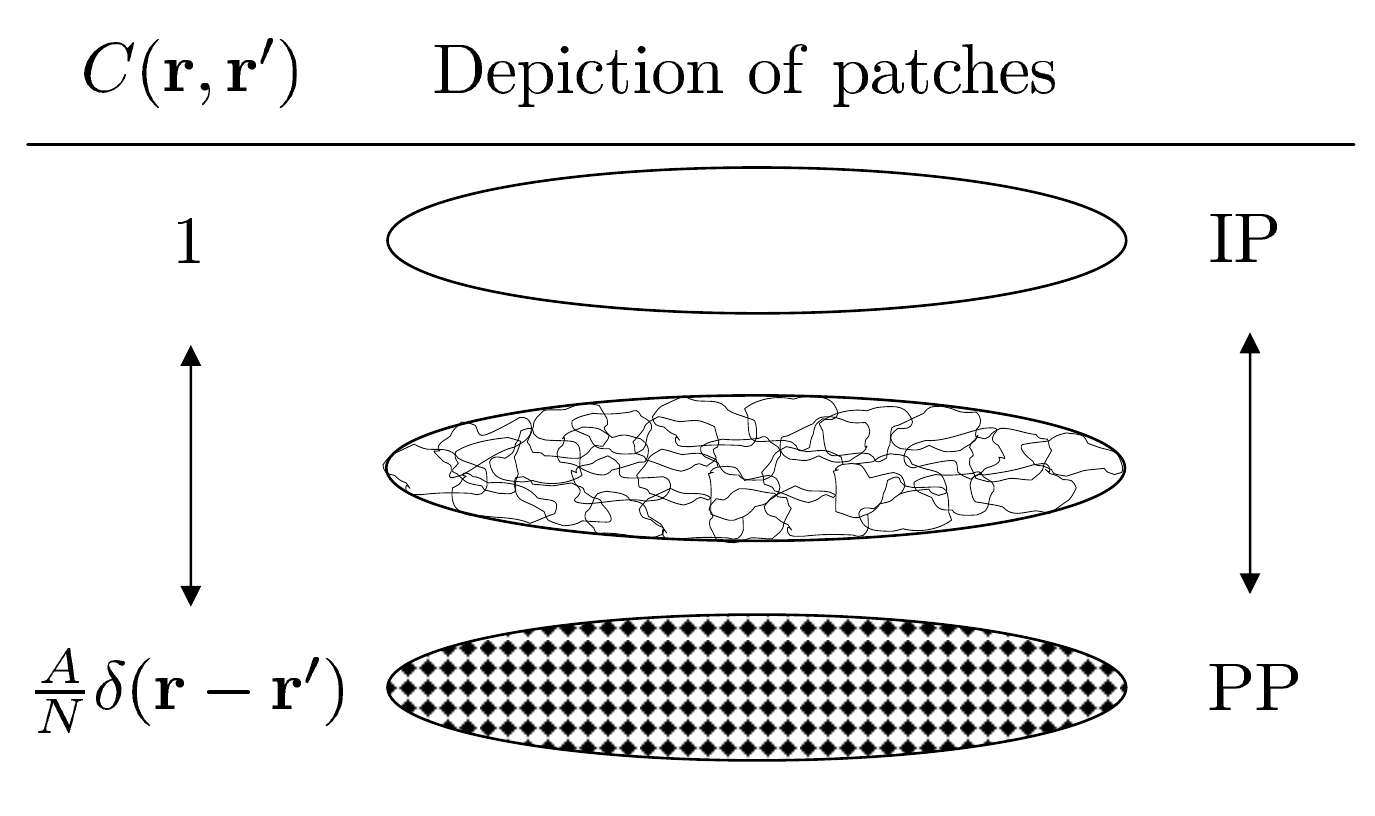}
\caption{Patch sizes and associated correlation functions, ranging from the Infinite Patch (IP) to the Point Patch (PP) from top to bottom.
\label{fig:patch_sizes}}
\end{figure}
Our second approach is motivated by the fact that, even in the IP and PP limits, it is often not the case that the geometric factor can be evaluated analytically. Furthermore, it is useful to evaluate the geometric factor for some specific correlation function that parameterize patch sizes. To accommodate with these issues, we present a formalism for the geometric factor that is tractable by numerical methods. We start with the eigenfunction expansion of the Green's and surface Green's functions~\cite{Jackson1999}:
\begin{eqnarray}\label{eq:greens_eigenfunction}
G(\mathbf{r},\mathbf{r}') &=& \sum_{i}\frac{f_{i}(\mathbf{r})f^{*}_{i}(\mathbf{r}')}{\lambda_{i}},\\
G_{\sigma}(\mathbf{r},\mathbf{r}') &=& -\sum_{i}\frac{f_{i}(\mathbf{r})\partial_{n'}f^{*}_{i}(\mathbf{r}')|_{\mathbf{r}'\in\sigma}}{\lambda_{i}},
\end{eqnarray}
where $f_{i}(\mathbf{r})$ are eigenfunctions of Laplace's equation with eigenvalues $\lambda_{i}$ and satisfy homogenous boundary conditions on the surface $\sigma$. Substituting the result into the expression for the geometric factor and using $G(\mathbf{r},\mathbf{r}')=G^{*}(\mathbf{r},\mathbf{r}')$ gives
\begin{eqnarray}\label{eq:geometricfactor_expansion}
\Lambda_{k}(\mathbf{r}) &=& \sum_{i,j}c_{ij}\nabla_{k}f_{i}(\mathbf{r})\nabla_{k}f^{*}_{j}(\mathbf{r})\nonumber,\\
c_{ij}&=&\frac{\int_{\sigma'}\int_{\sigma}C(\mathbf{r},\mathbf{r}')\partial_{n}f_{i}(\mathbf{r})\partial_{n'}f^{*}_{j}(\mathbf{r}')d\mathbf{r}d\mathbf{r}'}{\lambda_{i}\lambda^{*}_{j}}.
\end{eqnarray}
The geometric factor is now expressed as a double sum over eigenfunctions with expansion coefficients $c_{ij}$, which represent projections of the correlation function onto eigenfunctions of the surface geometry. The average patch size can then be modeled by specifying a correlation function that parameterizes the  patch size $\zeta$ (Fig.~\ref{fig:patch_sizes}), and projecting onto these eigenfunctions. 
The expansion coefficients $c_{ij}$ in the IP and PP limits are
\begin{equation}\label{eq:geometricfactor_expansion_coefficients}
c_{ij} = \frac{1}{\lambda_{i}\lambda^{*}_{j}}\times\begin{cases}
\int_{\sigma}\partial_{n}f_{i}(\mathbf{r})d\mathbf{r}\int_{\sigma}\partial_{n}f^{*}_{j}(\mathbf{r})d\mathbf{r}  &\text{IP,} \\
 \frac{A}{N}\int_{\sigma}\partial_{n}f_{i}(\mathbf{r})\partial_{n}f^{*}_{j}(\mathbf{r})d\mathbf{r} &\text{PP.}\end{cases}
\end{equation}

The patch correlation functions in Eq.~\ref{eq:CorrelationFunctionIPPP} represent two opposite extremes; however, Eq.~\ref{eq:geometricfactor_expansion} facilitates the study of the intermediate regime. Moreover, the latter raises the possibility of constructing more complex correlation functions out of a relatively small set of basis correlation functions that are simultaneously diagonalizable in the eigenfunction basis. This approach lends itself well to numerical modeling and is generally applicable to arbitrary geometries and patch correlation functions. By contrast, e.g., the choice of $C(\mathbf{r}, \mathbf{r}') \propto e^{-|\mathbf{r}-\mathbf{r}'|/\zeta}$ in Ref.~\cite{Dubessy2009} facilitates an analytical solution for an infinite planar surface, but not for other geometries. 

Patch sizes can be very dependent on material and surface preparation; however, values in the range $\sim10$~nm to $\sim1~\mu$m are typically reported by experiments~\cite{Camp1991,Camp1992,Speake2003}. It follows that, at ion-surface distances of $d\sim100~\mu$m that are typical to many ion trap experiments, the exact form of the correlation function will not influence the evaluation of electric field noise. We shall thus not concern ourselves with the form of the correlation function, nor with the physics governing it, and only use a simple parameterization of intermediate patch sizes by truncating the eigenfunction expansion beyond some higher-order term in the sum of Eq.~\ref{eq:geometricfactor_expansion}. Explicit examples of this method are given in section~\ref{sec:Spherical} and \ref{sec:Spheroidal}. 

\subsection{Scaling with surface distance} \label{sec:Model_scaling}
The focus of this work is on the scaling of the power spectrum of the electric field noise with surface distance, described through the geometric factor $\Lambda_{k}(\mathbf{r})$. Prior work on this subject has typically found this to be described by some power law with respect to surface distance $d$, i.e., $\Lambda_{k}(\mathbf{r})\propto d^{-\alpha}$ . We shall assume such a relationship to hold in general, and our parameter of interest, the scaling exponent $\alpha$, is evaluated as
\begin{equation}\label{eq:alpha}
\alpha_{k}(d) = - \partial_{\ln [d]} \left(\ln \left[\Lambda_{k}(\mathbf{r}_{0} + d\hat{d})\right]\right),
\end{equation}
where $\mathbf{r}_{0}$ is a reference point on the surface against which scaling in some direction $\hat{d}$ is evaluated. In the following we consider various surface geometries and we quantify their influence on the geometric factor through $\alpha$ in this way. Note that if $\Lambda_{k}(\mathbf{r})$ scales differently for the different $k$ modes, it is possible to obtain different values for $\alpha$ if, for example, motion normal instead of transverse to the electrode surface is considered. While this has been considered in past treatments of Johnson noise~\cite{Leibrandt2007}, it has largely been overlooked by patch potential models.


\section{Electrode geometry dependence}\label{sec:GeoDep}
We now proceed to consider the electrode geometry of the ion trap. The focus is on generic structures, including finite geometries, often found in experiments. We will consider planar electrodes and special cases thereof with hole-in-conductor, as well as spheroidal shapes such as spheres and very prolate spheroids as models for needles.

\subsection{Infinite planar electrode}
We first establish a connection with previous work in the field~\cite{Dubessy2009} by considering the infinite planar electrode. While inherently unphysical, this represents a good approximation to many ion trap geometries where the ion is confined only $\sim 100~\mu$m above a surface extending several mm in both directions of the plane~\cite{Chiaverini2005a}. It is also an excellent model for the geometry encountered in non-contact friction measurements with cantilevers~\cite{Stipe2001} -- a fact exploited by Dubessy et al.~\cite{Dubessy2009} to connect measurements of noise in these seemingly disparate systems through a single physical model.

\begin{figure}
\includegraphics[width=0.8\columnwidth]{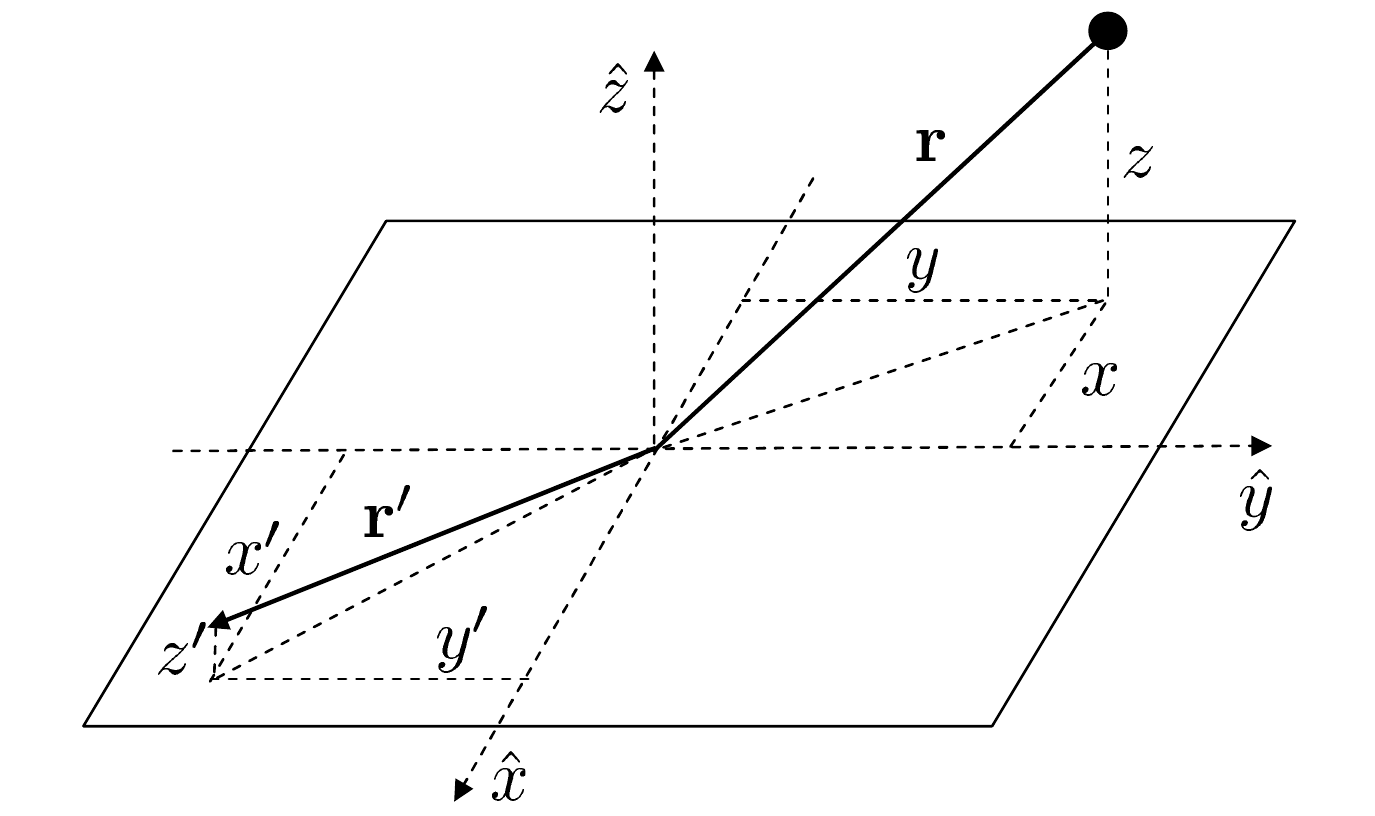}
\caption{Coordinate system used in the Green's function for the infinite plane in Eq.~\ref{eq:GInfPlane}. The black dot represents the ion.\label{fig:Greens_function_plane}}
\end{figure}

The Green's function for an infinite conductor is obtained from a straightforward consideration of image charges. In cartesian coordinates $(x,y,z)$ where the $\hat{z}$ axis is normal to the surface (Fig.~\ref{fig:Greens_function_plane}):
\begin{equation}\label{eq:GInfPlane}
G(\mathbf{r},\mathbf{r}')=\frac{1}{4\pi}\left[\frac{1}{\left|\mathbf{r}-\mathbf{r}'\right|}-\frac{1}{\left|\mathbf{r}-\mathbf{r}'+2z'\hat{z}\right|}\right].
\end{equation}

When an adsorbate with charge q at $\mathbf{r}'$ is brought much closer to the surface than the ion-surface distance, the corresponding potential seen by the ion is indistinguishable from that of a point patch potential $\phi = q z' \delta^{2}((x'-x)\hat{x}+(y'-y)\hat{y})$, which is consistent with earlier statement that the electric potential of a charge very close to a conducting surface is indistinguishable from an effective patch potential. The physical origin of $V_{i}(t)$ in this case would then be fluctuations in the charge-surface distance $z'$.

The infinite (point) patch limit is reached when the ion-surface distance is much smaller (larger) than the average patch size. By substituting the Green's function of Eq.~\ref{eq:GInfPlane} into the expression for $\Lambda_{k}(\mathbf{r})$ in Eq.~\ref{eq:geometricfactor}, we obtain
\begin{eqnarray}
\Lambda_{z}(d\hat{z}) &=& \begin{cases}0 &\text{IP,} \\ \frac{A}{N} \frac{3}{16\pi} d^{-4} &\text{PP,}\end{cases}\nonumber \\
\Lambda_{x,y}(d\hat{z}) &=& \begin{cases}0 &\text{IP,} \\ \frac{A}{N} \frac{3}{32\pi}d^{-4} &\text{PP.}\end{cases}
\end{eqnarray}

Ion trap experiments studying this geometry, such as those based on the surface electrode ion trap [Fig.~\ref{fig:planar_trap} (d)], have focused on the point patch limit as ion heights of $d\sim100~\mu$m and a trap size of $\sim10$~mm, respectively, are typical. At these dimensions, patch sizes of $\sim1~\mu$m appear as point patches and the model thus predicts an $\alpha=4$ scaling. As the ion-surface distance is lowered, one expects $\alpha$ to decrease, as was also concluded by Dubessy et al.~\cite{Dubessy2009}. In principle, a continuous transition from $4$ up to $0$ is predicted by our model; however, we note that the IP limit is trivial, as an infinite plane at uniform potential has no electric field. Furthermore, it is clear that, as systems are scaled to small surface distances below the dimensions of the patches, the exact form of the patch correlation function will have a strong impact on the geometric factor~\cite{Dubessy2009}.

\subsection{Infinite planar electrode with a hole}
\begin{figure}
\includegraphics[width=0.8\columnwidth]{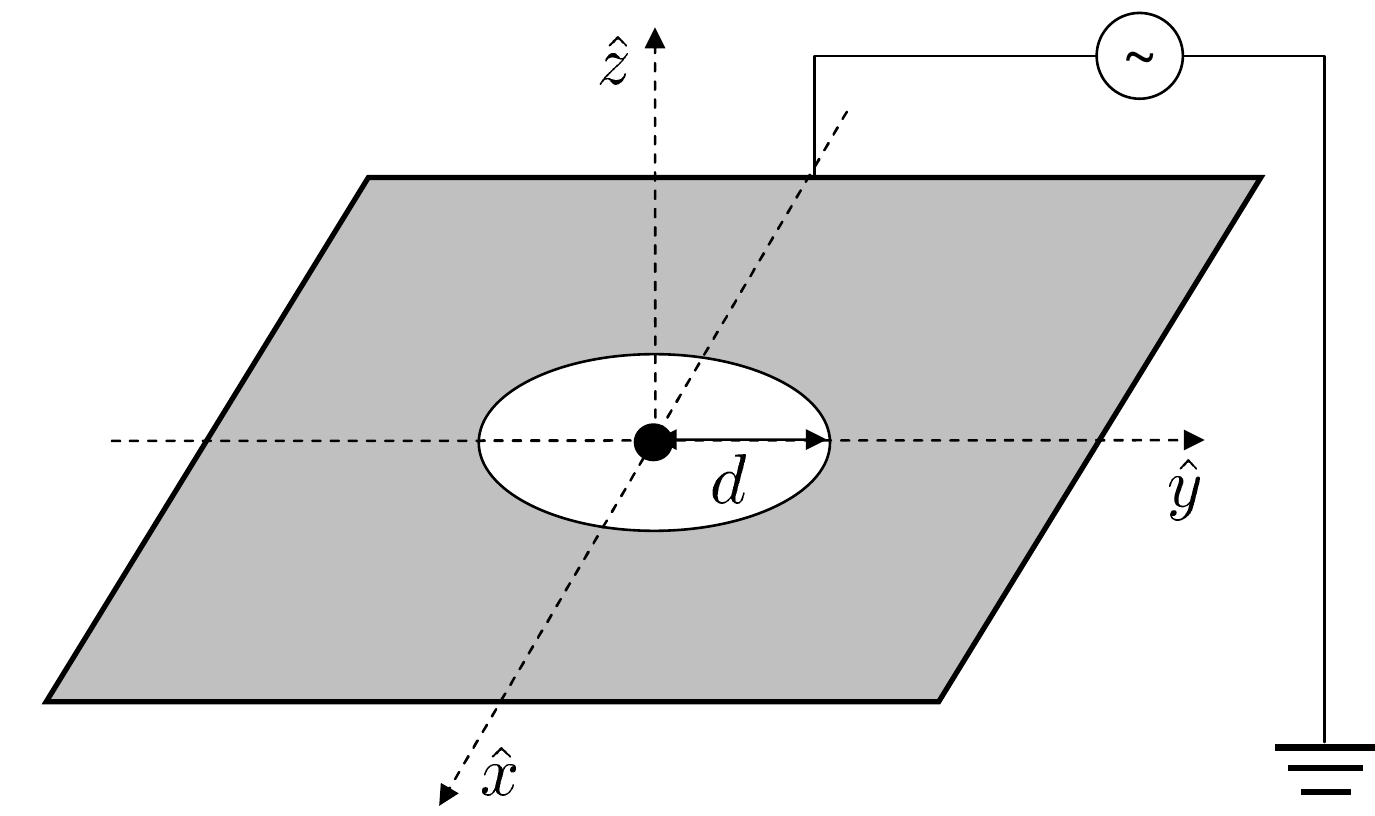}
\caption{Ion (black dot) in a planar hole trap with radius $d$. RF is applied to the entire plane.
\label{fig:planar_hole_trap}}
\end{figure}

\begin{figure}
\includegraphics[width=0.8\columnwidth]{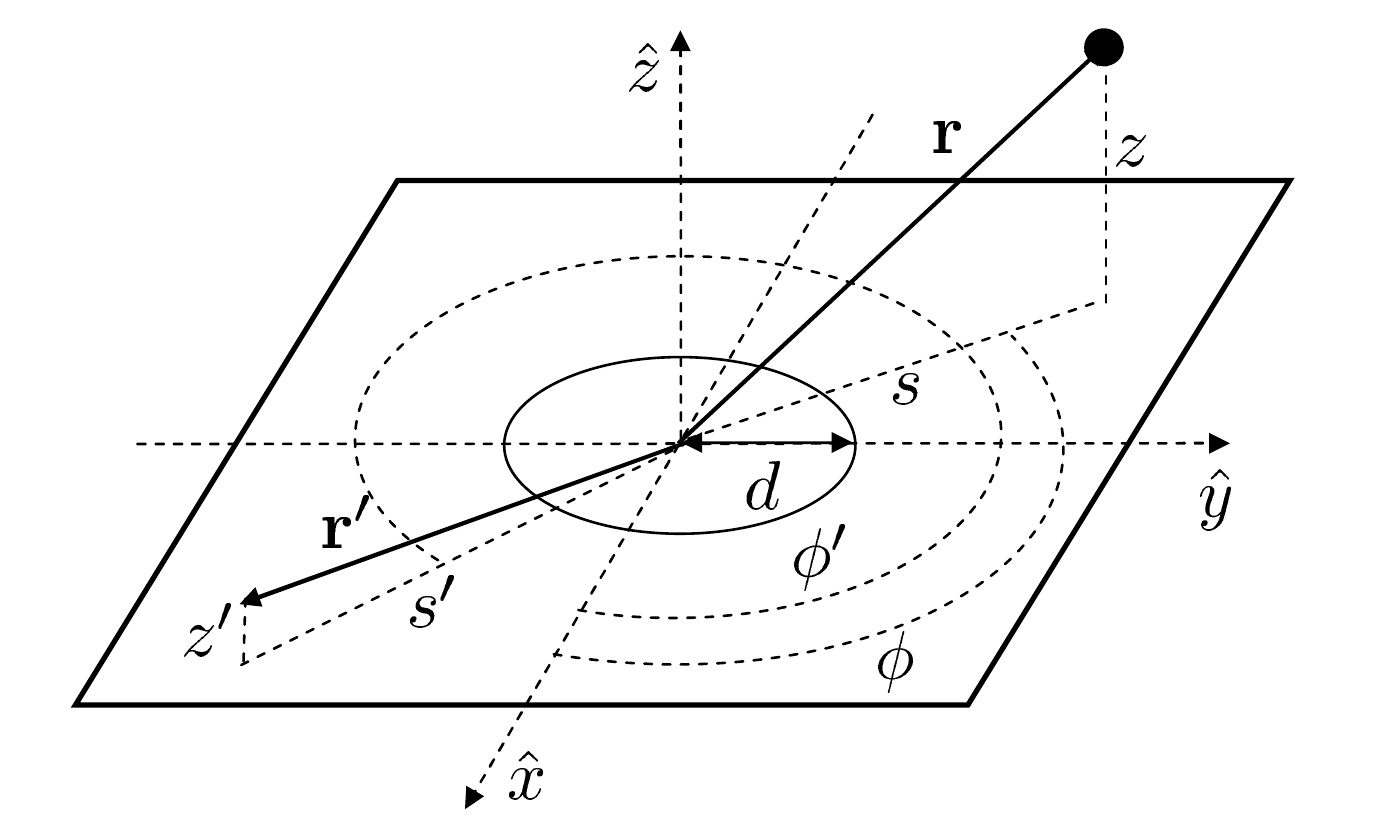}
\caption{Coordinates system used for the hole trap. The ion (black dot) is suspended at point $\mathbf{r}$. The source is located at $\mathbf{r}'$.
\label{fig:greens_function_hole}}
\end{figure}

This section considers the geometry of the hole trap studied previously in e.g. Refs.~\cite{Brewer1992,DeVoe2002} and depicted in Fig.~\ref{fig:planar_hole_trap}. This geometry is also of relevance to large-scale quantum information processing where an architecture based on an array of such traps has been proposed ~\cite{Cirac2000}.
We model this hole trap as a thin, infinite conducting sheet with a circular hole. As the ion is always trapped at the center of the hole, the relevant scaling parameter becomes the hole radius, which we label here as $d$. In cylindrical coordinates (Fig.~\ref{fig:greens_function_hole}), where
\begin{equation}
x = s \cos{\phi},\quad y = s \sin{\phi},\quad z = z,
\end{equation}
the Green's function for a thin conductor with a hole is~\cite{Eberlein2011}
\begin{eqnarray}\label{eq:greens_function_hole}
G(\mathbf{r},\mathbf{r}') &=& \frac{1}{8\pi} \left\{\frac{1}{\Gamma_{-}}\left[1+\frac{2}{\pi}\arctan{\left(\frac{\Psi_{-}}{\Gamma_{-}}\right)} \right] \right. \nonumber \\
&&\left.-\frac{1}{\Gamma_{+}}\left[1+\epsilon\frac{2}{\pi}\arctan{\left(\frac{\Psi_{+}}{\Gamma_{+}}\right)} \right] \right\},
\end{eqnarray}
where $\mathbf{r},\mathbf{r}'$ lie on the same side of the plane, and where
\begin{equation}
\Gamma_{\mp}=\sqrt{s^{2}+s'^{2}-2ss' \cos{\left(\phi-\phi'\right)}+\left(z\mp z'\right)^{2}},
\end{equation}
\begin{eqnarray}
\Psi_{\mp} &=& \frac{1}{\sqrt{2}d} \left\{\left(s^{2}+z^{2}-d^{2}\right)\left(s'^{2}+z'^{2}-d^{2}\right)\right.\nonumber\\
&&\pm 4d^{2}zz'+\sqrt{\left[z^{2}+\left(s-d\right)^{2}\right]\left[z^{2}+\left(s+d\right)^{2}\right]}\nonumber\\
&&\times  \left.\sqrt{\left[z'^{2}+\left(s'-d\right)^{2}\right]\left[z'^{2}+\left(s'+d\right)^{2}\right]}\right\}^{1/2}
\end{eqnarray}
and
\begin{equation}
\epsilon=\text{sgn}\left[z\left(s'^{2}+z'^{2}-d^{2}\right)+z'\left(s^{2}+z^{2}-d^{2}\right)\right].
\end{equation}

We place the ion at the center of the hole at $\mathbf{r}=(0,0,0)$ and substitute the Green's function for a conductor with a hole (Eq.~\ref{eq:greens_function_hole}) into the geometric factor (Eq.~\ref{eq:geometricfactor}). Care must be taken when evaluating the integral in the geometric factor as the sharp edge of the hole at $s = d$ results in a divergence. To circumvent this issue, one may introduce a small parameter $\delta \ll d$ and perform the radial integration over $(d+\delta,\infty)$. Doing so we obtain
\begin{eqnarray}
\Lambda_{z}(0) &=& \begin{cases}\frac{1}{4}d^{-2} &\text{IP,} \\  \frac{A}{N}\frac{1}{32\pi} d^{-4} &\text{PP,}\end{cases}\nonumber \\
\Lambda_{s}(0) &=&\begin{cases}0 &\text{IP,} \\  \frac{A}{N}\frac{1}{4\pi^{3}}\left(2 \ln\frac{d}{2\delta}-3)\right) d^{-4} & \text{PP.}\end{cases}
\end{eqnarray}
Note that the logarithmic divergence in $d/\delta$ for radial scaling does not detract from the main result of $\alpha = 4$.

Unlike the infinite plane, the IP limit of $\Lambda_{z}(0)$ does not evaluate to zero as only one side of the sheet is held at a non-zero potential. It is furthermore interesting to note that similar results for the scaling of the geometric factor are obtained as for the infinite plane despite the significant difference between the two geometries.

\subsection{Spherical electrodes} \label{sec:Spherical}
\begin{figure}
\includegraphics[width=0.8\columnwidth]{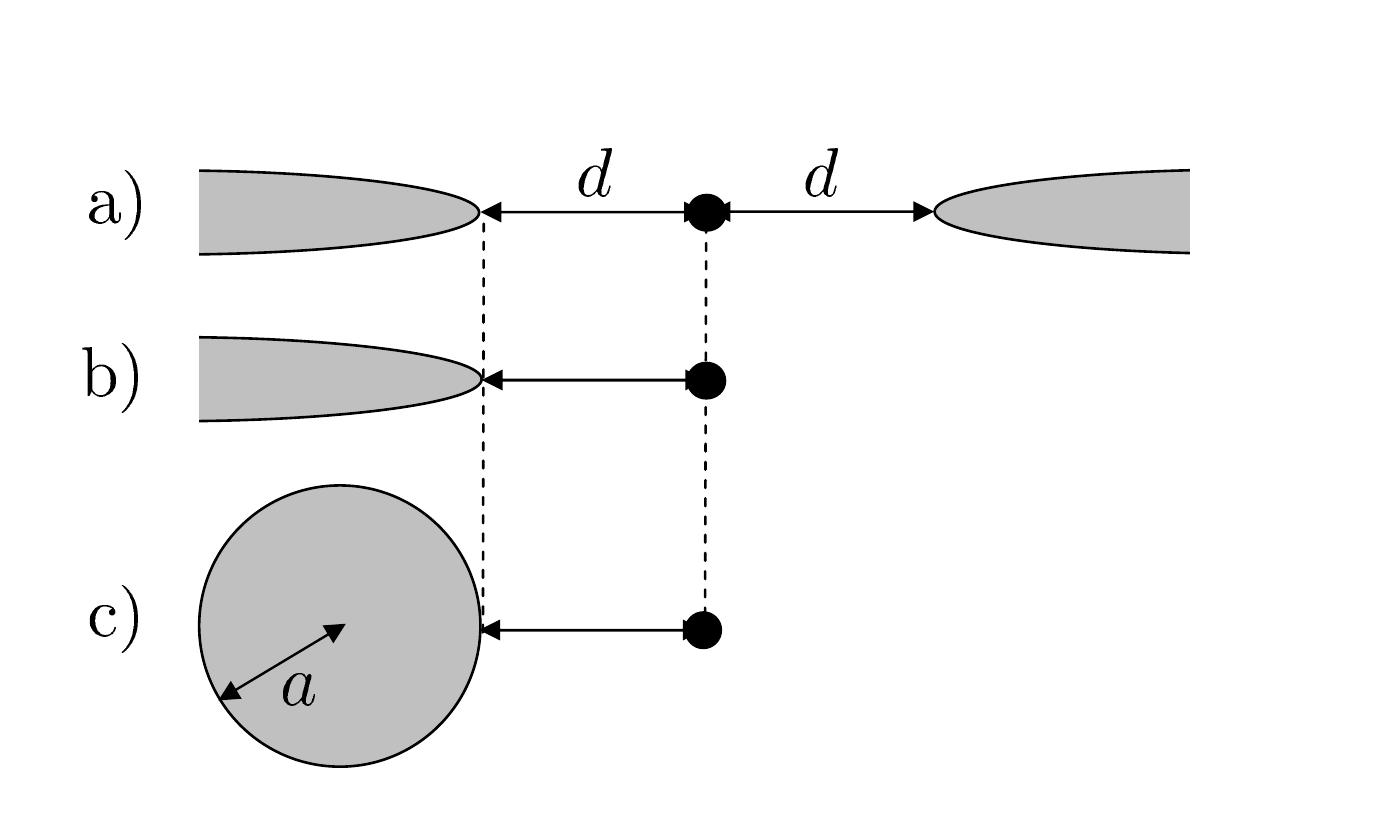}
\caption{Approximations to the Deslauriers et al. needle trap~\cite{Deslauriers2006}. Black dot: ion. Shaded electrode: RF. a) Needle trap configuration. b) Approximation to a) assuming a single needle suffices to describe the scaling of the geometric factor with respect to $d$. c) approximation to b) representing the needle by an effective sphere of radius $a$. \label{fig:needle_bispherical_sphere}}
\end{figure}

We now consider a spherical electrode, which is an example of a finite geometry where placing an ion outside the sphere and far from its surface is well defined as well as an example of a geometry radically different from the two-dimensional planar surface. Nevertheless, for small surface distances, one expects to recover the results for the two-dimensional scenario. The geometry furthermore represents a reasonable approximation to the tip of the needles of the Deslauriers et al. needle trap~\cite{Deslauriers2006}. This is illustrated in Fig.~\ref{fig:needle_bispherical_sphere}, where the three images from top to bottom represent the successive approximations made in going from the real experiment configuration to the geometry modeled here and in the following section.

\begin{figure}
\includegraphics[width=0.8\columnwidth]{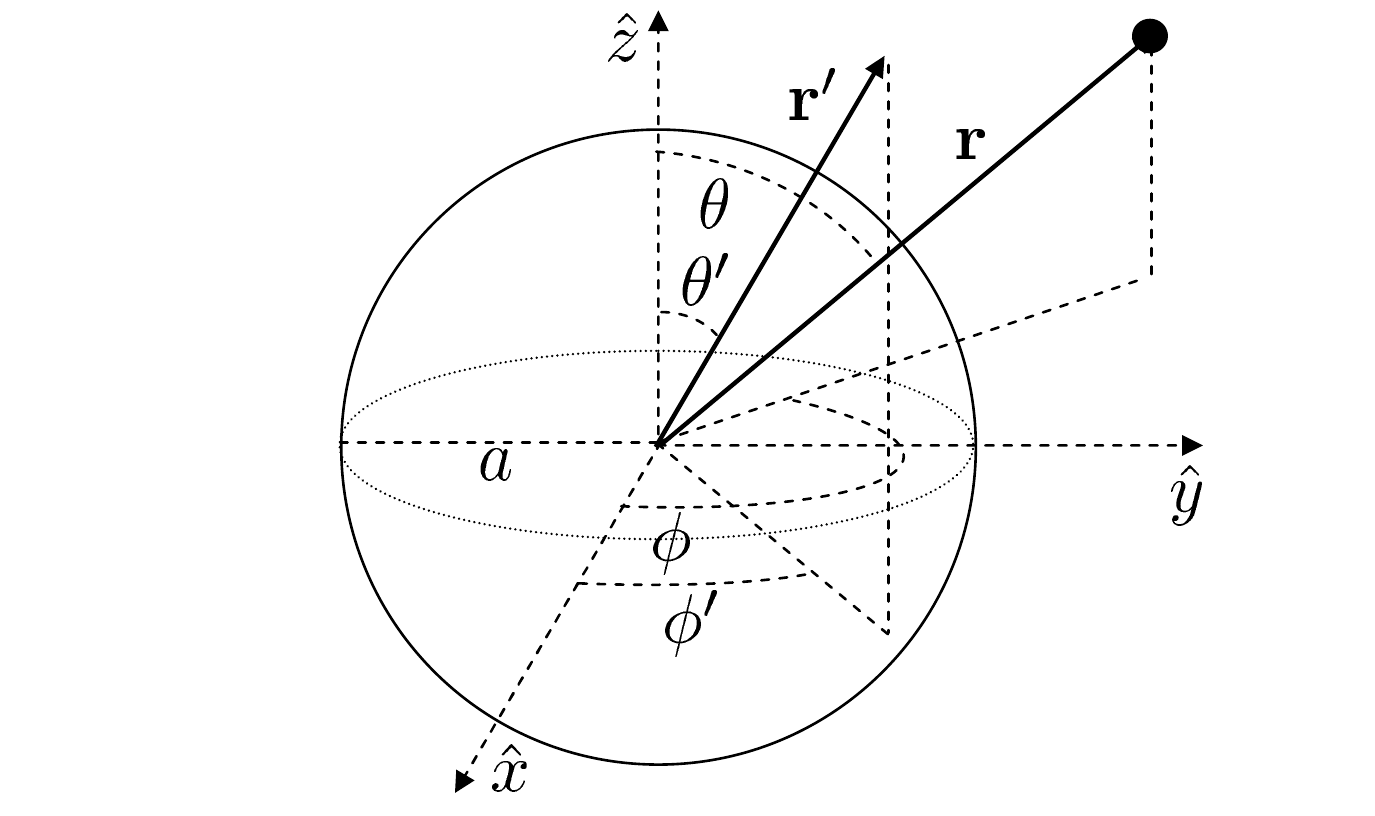}
\caption{Coordinates system used for the sphere. The ion is suspended at point $\mathbf{r}$. The source is located at $\mathbf{r}'$
\label{fig:greens_function_sphere}}
\end{figure}
The Green's function for this system expressed in spherical coordinates defined in Fig.~\ref{fig:greens_function_sphere} is
\begin{equation}
G (\mathbf{r},\mathbf{r'}) = \frac{1}{4\pi}\left[\frac{1}{\left|\mathbf{r}-\mathbf{r}'\right|} - \frac{1}{\frac{r'}{a}\left|\mathbf{r}-\frac{a^{2}}{r'^{2}}\mathbf{r}'\right|}\right].
\end{equation}
Substituting into Eq.~\ref{eq:geometricfactor}, we obtain the following limiting forms for the geometric factor for an ion at distance $r$ from the center of a unit sphere ($a=1$):
\begin{eqnarray}\label{eq:SEomega5}
\Lambda_{r}(\mathbf{r})&=&\begin{cases}r^{-4}  &\text{IP} \\ \frac{A}{N}\frac{3+8 r^{2} + r^{4}}{4\pi\left(r^{2}-1\right)^{4}} &\text{PP}\end{cases}, \nonumber \\
\Lambda_{\theta}(\mathbf{r})&=&\begin{cases}0  &\text{IP} \\ \frac{A}{N}\frac{3(1+ r^{2})}{4\pi\left(r^{2}-1\right)^{4}} &\text{PP}\end{cases}.
\end{eqnarray}
Unlike the planar geometries, the geometric factor now has a non-trivial variation with distance.

Although this geometry allows for a derivation of $\alpha$ in a closed-form, it is extremely helpful to consider the eigenfunction expansion of the geometric factor (Eq.~\ref{eq:geometricfactor_expansion}). This approach lets us evaluate the geometric factor for intermediate patch sizes between the IP and PP limits (Eq.~\ref{eq:geometricfactor_expansion_coefficients}) from which we may gain intuition about how patch size affects the scaling of the power spectrum of electric field noise. Invoking this approach for a geometry that provides for a closed-form solution furthermore provides for a consistency check between the two approaches.

One could treat this problem rigorously by first solving Helmholtz's equation to obtain the eigenfunctions, but it is more expedient to start from the surface Green's function, which assumes the simple form:
\begin{equation}\label{eq:sphere_greens_function}
G_{\sigma} (\mathbf{r},\mathbf{r'})= \sum^{\infty}_{l=0}\sum^{l}_{m=-l}\frac{a^{l+1}}{r^{l+1}}Y_{lm}(\theta',\phi')Y^{*}_{lm}(\theta,\phi).
\end{equation}
Here the $Y_{lm}$ are the spherical harmonics of degree $l$ and order $m$ normalized such that $\oint Y_{lm}(\theta,\phi)Y^{*}_{l'm'}(\theta,\phi)d\Omega = \delta_{ll'}\delta_{mm'}$. Substituting into Eq.~\ref{eq:geometricfactor},
\begin{eqnarray}\label{eq:SEomega5transverse}
\Lambda_{r}(\mathbf{r})&=&\begin{cases}4\pi\left|\partial_{r}\frac{Y_{00}(\theta,\phi)}{r}\right|^{2}  &\text{IP} \\\frac{A}{N} \sum^{\infty}_{l=0}\sum^{l}_{m=-l}\left|\partial_{r}\frac{Y_{lm}(\theta,\phi)}{r^{l+1}}\right|^{2}&\text{PP}\end{cases}, \nonumber \\
\Lambda_{\theta}(\mathbf{r})&=&\begin{cases}0  &\text{IP} \\ \frac{A}{N}\sum^{\infty}_{l=0}\sum^{l}_{m=-l}\left|\partial_{\theta}\frac{Y_{lm}(\theta,\phi)}{r^{l+2}}\right|^{2}&\text{PP}\end{cases}.
\end{eqnarray}

An advantage of the eigenfunction expansion of the geometric factor in terms of spherical harmonics is that one may truncate higher order terms beyond some $l=l_{0}\geq0$. Such an action roughly corresponds to evaluating the geometric factor for patches of angular radii $\theta_{\zeta}\sim2/l_{0}$ for large $l_{0}\gg0$. Instead of sharp truncation, patch size dependance could also be introduced by attenuating higher order terms according to some distribution, resulting in different scaling behavior close to the surface. 

\begin{figure}
\includegraphics[width=0.9\columnwidth]{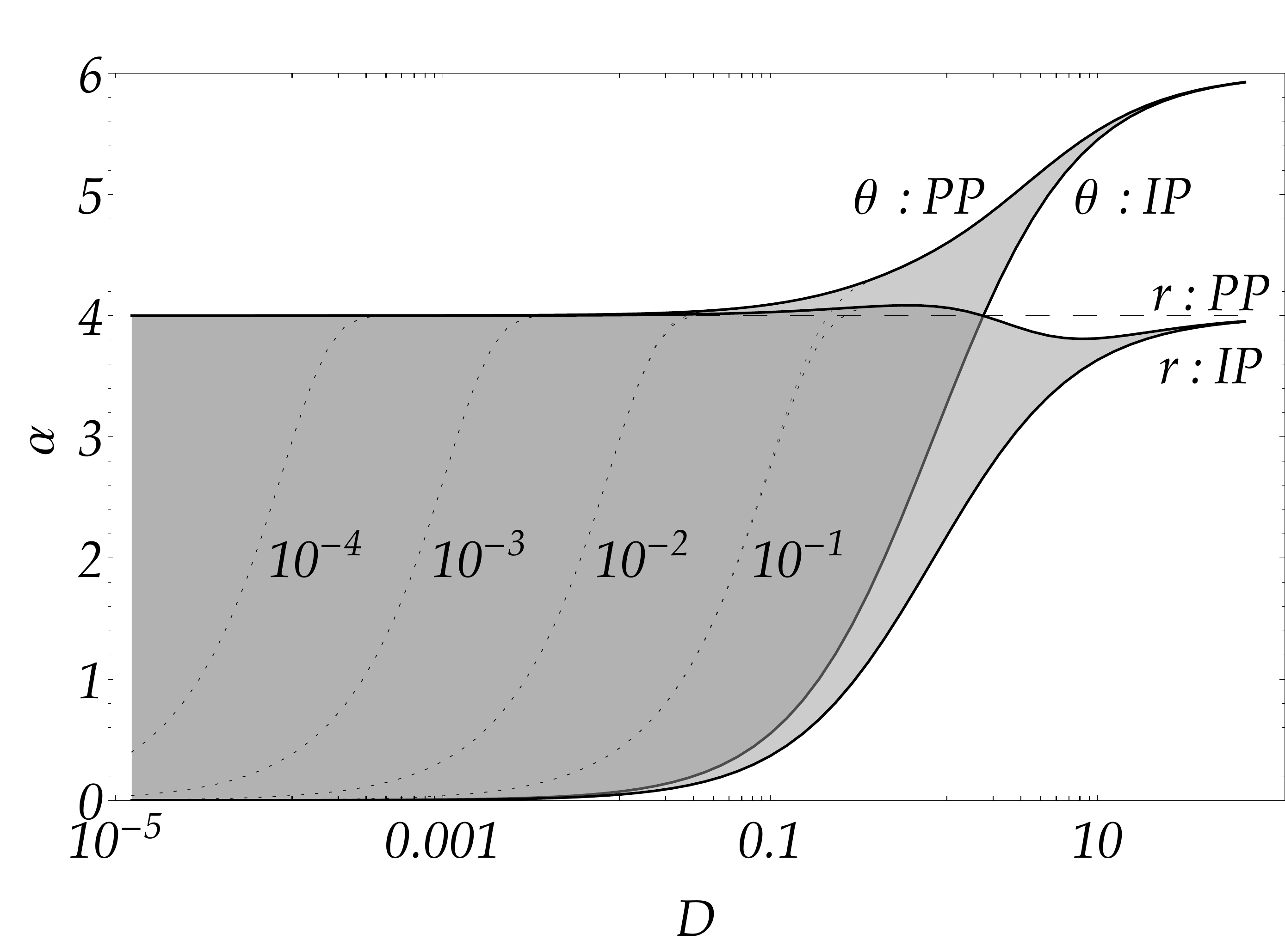}
\caption{Scaling exponent $\alpha$ as a function of dimensionless distance $D=d/a$ above a sphere of radius $a$ for both normal ($r$) and transverse ($\theta$) modes. The labeled solid lines are for the IP and PP limits, and are based on Eq.~\ref{eq:SEomega5}. The shaded area between the IP and PP lines of each mode represents permissible values of $\alpha$. The dotted lines correspond to intermediate patches sizes $\theta_{\zeta}/a=10^{-1},10^{-2},10^{-3},10^{-4}$ evaluated via Eq.~\ref{eq:SEomega5transverse} by truncating the expansion at $l_{0}=2/(\theta_{\zeta})$. The dashed line of $\alpha=4$ is plotted for reference.
\label{fig:GLOW_SAFEF_sphere}}
\end{figure}

The scaling exponent $\alpha$ is evaluated, using both approaches presented above, with respect to dimensionless distance $D=d/a$ such that all dimensions are in units of electrode dimension, $a$. We make use of Eq.~\ref{eq:alpha} to extract $\alpha$ from the geometric factor and we perform the evaluation with respect to the surface of the unit sphere for both $(r,\theta)$ modes by setting $\mathbf{r}_{0}= \hat{d}=\hat{r}$. Additionally, we exploit the spherical symmetry to set $\theta=\phi=0$. Figure~\ref{fig:GLOW_SAFEF_sphere} shows the scaling for the radial, $r$, mode as well as the transverse, $\theta$, mode.

The predictions in the limiting cases, given by Eq.~\ref{eq:SEomega5}, are drawn in Fig.~\ref{fig:GLOW_SAFEF_sphere} as thick, solid lines, with IP and PP being the upper and lower lines, respectively. Similarly, the results of the approach based on the eigenfunction expansion, given in Eq.~\ref{eq:SEomega5transverse}, are drawn as thin, dotted lines for patch sizes of $\theta_{\zeta}$ ranging from $10^{-4}-10^{-1}$ for the $r$ and $\theta$ modes as annotated in the figure. Note that for the $\theta$ mode, we have omitted the trivial case of truncating at $l_{0}=0$, corresponding to a pure monopole with no component in the $\theta$ direction.

As expected, the predictions of the eigenfunction expansion are all bounded by the solid lines representing the limiting cases of IP and PP. In the limit of $D\ll\theta_{\zeta}$, the results of the eigenfunction expansion converge towards the IP limit, since at infinitesimal distances, the finite patches appear infinite. The scale at which this occurs, however, is obviously patch size dependent. For example, if $a=1$~cm, realistic patch sizes of $1~\mu$m are obtained with $\theta_{\zeta}=10^{-4}$ and the crossover between the IP and PP limits occur for surface distances of order $1~\mu$m. This trend of decreasing $\alpha$ with decreasing distance scale $D$ is in agreement with the infinite plane model of Dubessy et al.~\cite{Dubessy2009}. 

In the limit of $D\gg 1$, the scaling for all patch sizes converges toward $4$ for the $r$ mode and $6$ for the $\theta$ mode and become independent of the exact patch size, since at large distances, the ion is no longer sensitive to the structure of the surface. The result for the $\theta$ mode is not obtainable in the infinite plane approximation. It is also worth noting that for $D\sim10$ -- a scale commensurate with the Deslauriers et al. needle trap experiment ~\cite{Deslauriers2006} -- the scaling exponent for the $r$ mode is consistent with the $\alpha=3.5\pm 0.1$ value obtained by that experiment.

\subsection{Spheroidal electrodes} \label{sec:Spheroidal}
\begin{figure}
\includegraphics[width=0.9\columnwidth]{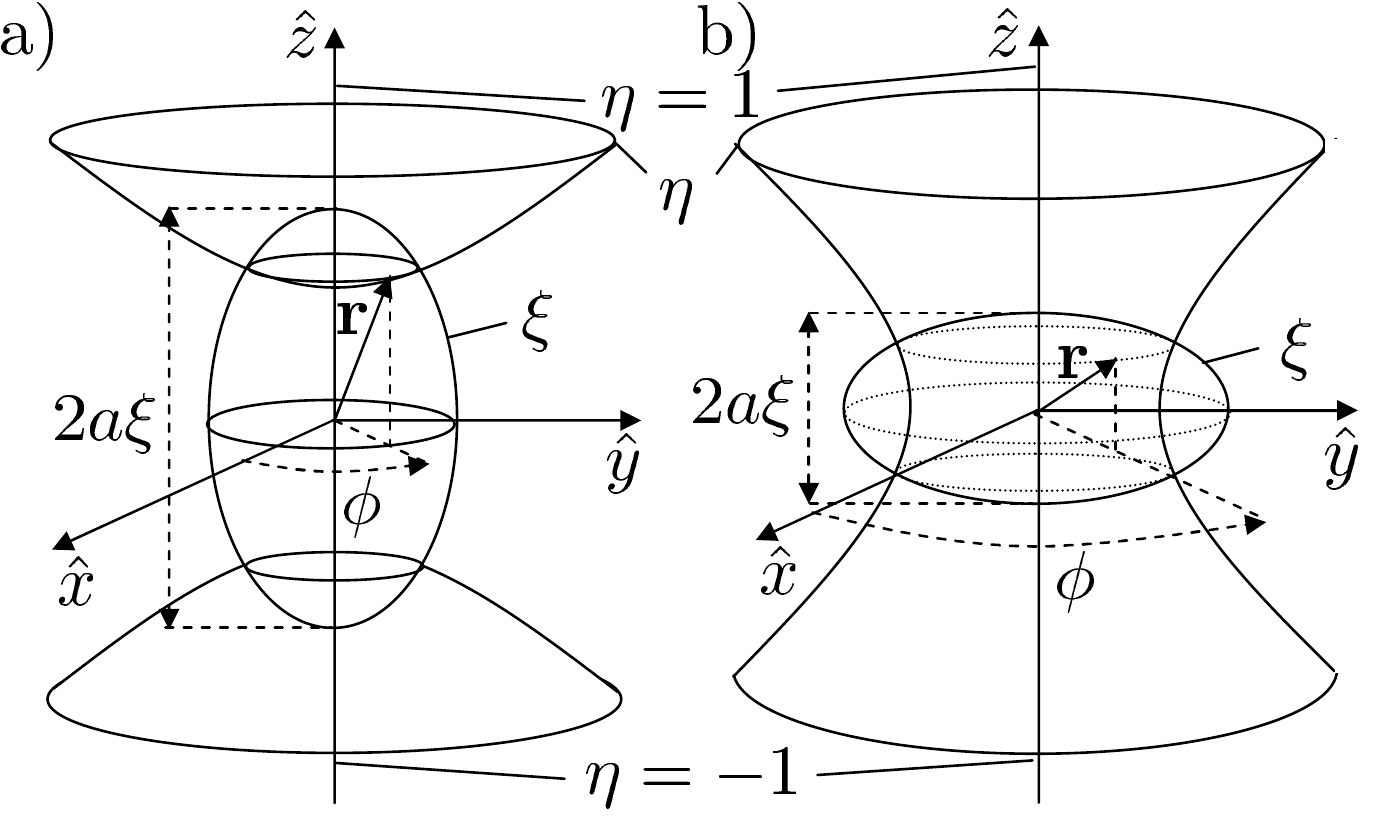}
\caption{(a) Prolate spheroidal coordinates defined in Eq.~\ref{eq:prolate_coor}. (b) oblate spheroidal coordinates defined in Eq.~\ref{eq:oblate_coor}.
 \label{fig:prolateoblate}}
\end{figure}

We now consider an ion suspended along the $\hat{z}$ axis outside electrodes of spheroidal geometries (Fig.~\ref{fig:prolateoblate}). The spheroids approximate a wide range of shapes, and generalize our discussion of spheres in spherical coordinates above. For example, a thin prolate spheroid in place of a sphere (Fig.~\ref{fig:needle_bispherical_sphere}), would be a more reasonable approximation for the needle trap described in Ref.~\cite{Deslauriers2006}. On the other hand, a finite planar trap, or disc, could be described as an extremely flat oblate ellipsoid. We may collectively model these geometries by solving Laplace's equation in spheroidal coordinates, where the geometry is effectively specified via a choice of either prolate or oblate coordinate system.

We use the following definitions (c.f. Fig.~\ref{fig:prolateoblate}) for prolate spheroidal coordinates $(1\le\xi<\infty, -1\le\eta\le 1, 0\le\phi \le 2\pi)$
\begin{eqnarray}
\label{eq:prolate_coor}
x &=& a \sqrt{(\xi^{2}-1)(1-\eta^{2})}\cos{\phi}, \nonumber\\
y &=& a \sqrt{(\xi^{2}-1)(1-\eta^{2})}\sin{\phi},\nonumber\\
z &=& a \xi\eta,
\end{eqnarray}
and oblate spheroidal coordinates $(0\le\xi<\infty, -1\le\eta\le 1, 0\le\phi \le 2\pi)$
\begin{eqnarray}
\label{eq:oblate_coor}
x &=& a \sqrt{(1+\xi^{2})(1-\eta^{2})}\cos{\phi}, \nonumber\\
y &=& a \sqrt{(1+\xi^{2})(1-\eta^{2})}\sin{\phi},\nonumber\\
z &=& a \xi\eta.
\end{eqnarray}
In this notation, the $\eta$ and $\xi$ coordinates are the spheroidal analogues of the spherical coordinates $\cos{\theta}$ and $r$ respectively. For example, in prolate spheroidal coordinates, surfaces of constant $\xi=\xi_{0}$ form elongated ellipsoids. The degenerate limit $\xi_{0}\rightarrow 1$ is a straight line between $z = \pm a$, while the other limit $\xi_{0}\rightarrow \infty$ is a sphere. In oblate spheroidal coordinates, surfaces of constant $\xi=\xi_{0}$ form flattened ellipsoids, and the limits $\xi_{0}\rightarrow0$ and $\xi_{0}\rightarrow\infty$ correspond to a disc and a sphere, respectively. For simplicity, we consider the unit spheroid, meaning that all dimensions are in units of $a$.

Unlike in the analysis for planes and spheres, the Green's function for these geometries cannot be written in closed form. However, Laplace's equation is separable in the spheroidal coordinates, and the general solution may be expressed as a series of Legendre polynomials, $P_{lm}(x)$ and $Q_{lm}(x)$, of the first and second kind, respectively, defined with normalization
\begin{equation}
\int^{1}_{-1}P_{lm}(x)P_{l'm}(x)dx = \delta_{ll'}.
\end{equation}
Hence, similar to the eigenfunction expansion of the geometric factor, used in our treatment of spheres, we start from the surface Green's function for prolate and oblate coordinates. We then examine the scaling of the geometric factor $\Lambda_{k}(\mathbf{r})$ in Eq.~\ref{eq:geometricfactor} along the $\hat{z}$ axis, where $\eta=1$. Both at very small and at very large distances from the electrodes, we intuitively expect scaling laws similar to those found for the sphere; however, at intermediate distances, the shape of the electrodes is expected to become important.

Given an ellipsoid with surface $\xi_{0}$, the prolate spheroid surface Green's function is
\begin{equation}\label{eq:greens_function_prolate}
G_{\sigma}(\mathbf{r},\mathbf{r'})=\frac{\sum^{\infty}_{l,m}\frac{Q_{lm}(\xi)}{Q_{lm}(\xi_{0})}P_{lm}(\eta)P_{lm}(\eta')e^{im(\phi-\phi')}}{2\pi\sqrt{(\xi_{0}^{2}-1)(\xi_{0}^{2}-\eta'^{2})}}
\end{equation}
and the oblate spheroid surface Green's function is
\begin{equation}\label{eq:greens_function_oblate}
G_{\sigma}(\mathbf{r},\mathbf{r'})=\frac{\sum^{\infty}_{l,m}\frac{Q_{lm}(i\xi)}{Q_{lm}(i\xi_{0})}P_{lm}(\eta)P_{lm}(\eta')e^{im(\phi-\phi')}}
{2\pi\sqrt{(1+\xi_{0}^{2})(\xi_{0}^{2}+\eta'^{2})}},
\end{equation}
where $m=-l,-l+1,\cdots,l$ in the sum. By substituting Eq.~\ref{eq:greens_function_prolate} and \ref{eq:greens_function_oblate} into Eq.~\ref{eq:geometricfactor}, the geometric factors for prolate and oblate spheroids become
\begin{widetext}
\begin{equation}\label{eq:SEomegaProlate}
\Lambda_{k}^\mathrm{prolate}(\mathbf{r}) =
\begin{cases}
2\left|\frac{\nabla_{k} Q_{00}(\xi)}{Q_{00}(\xi_{0})}\right|^{2}
&\text{IP}
\\
\frac{\sum^{\infty}_{l,l',m}
c^{-}_{ll'm} \nabla_{k}f_{lm}(\xi,\eta,\phi,\xi_{0})
\nabla_{k}f^{*}_{l'm}(\xi,\eta,\phi,\xi_{0})}{\sqrt{\xi_{0}^{2}-1}}
&\text{PP}
\end{cases},
\end{equation}
\begin{equation}\label{eq:SEomegaOblate}
\Lambda_{k}^\mathrm{oblate}(\mathbf{r}) =
\begin{cases}
2\left|\frac{\nabla_{k} Q_{00}(i \xi)}{Q_{00}(i\xi_{0})}\right|^{2}
&\text{IP}
\\
\frac{\sum^{\infty}_{l,l',m}c^{+}_{ll'm} \nabla_{k}f_{lm}(i\xi,\eta,\phi,i\xi_{0})
\nabla_{k}f^{*}_{l'm}(i\xi,\eta,\phi,i\xi_{0})}{\sqrt{1+\xi_{0}^{2}}}
& \text{PP}
\end{cases},
\end{equation}
\end{widetext}
where the sum over $m$ is for $|m|\le\text{min}(l,l')$ and where
\begin{equation}
c^{\pm}_{ll'm}=\frac{A}{N}\int^{1}_{-1}\frac{P_{lm}(\eta)P_{l'm}(\eta)}{\sqrt{\xi_{0}^{2}\pm\eta^{2}}}d\eta,\\
\end{equation}
and
\begin{equation}
f_{lm}(\xi,\eta,\phi,\xi_{0})=\frac{Q_{lm}(\xi)}{Q_{lm}(\xi_{0} )}\frac{P_{lm}(\eta)e^{im\phi}}{\sqrt{2\pi}}.
\end{equation}
We note that these results reduce to those of the sphere in the limit $\xi_{0}\rightarrow\infty$ as expected, since $\lim_{\xi\rightarrow\infty}Q_{lm}(\xi) \propto \xi^{-l}$.

We evaluate the geometric factor numerically for the $\xi$ and $\eta$ modes along the $z$ axis. In this direction, all terms with $m\neq 0$ ($|m|\neq 1$) vanish for the $\xi$ $(\eta)$ mode. The evaluation is complicated by the cross terms $c^{\pm}_{ll'm}$ as the system is not invariant with respect to the $\eta$ coordinate. However, not all terms are significant as $c^{\pm}_{ll'm}$ is observed to go as $\sim e^{-|l-l'|\sqrt{\xi_{0}^{2}-1}/\xi_{0}}$ for large $|l-l'|$.

\begin{figure}
\includegraphics[width=0.9\columnwidth]{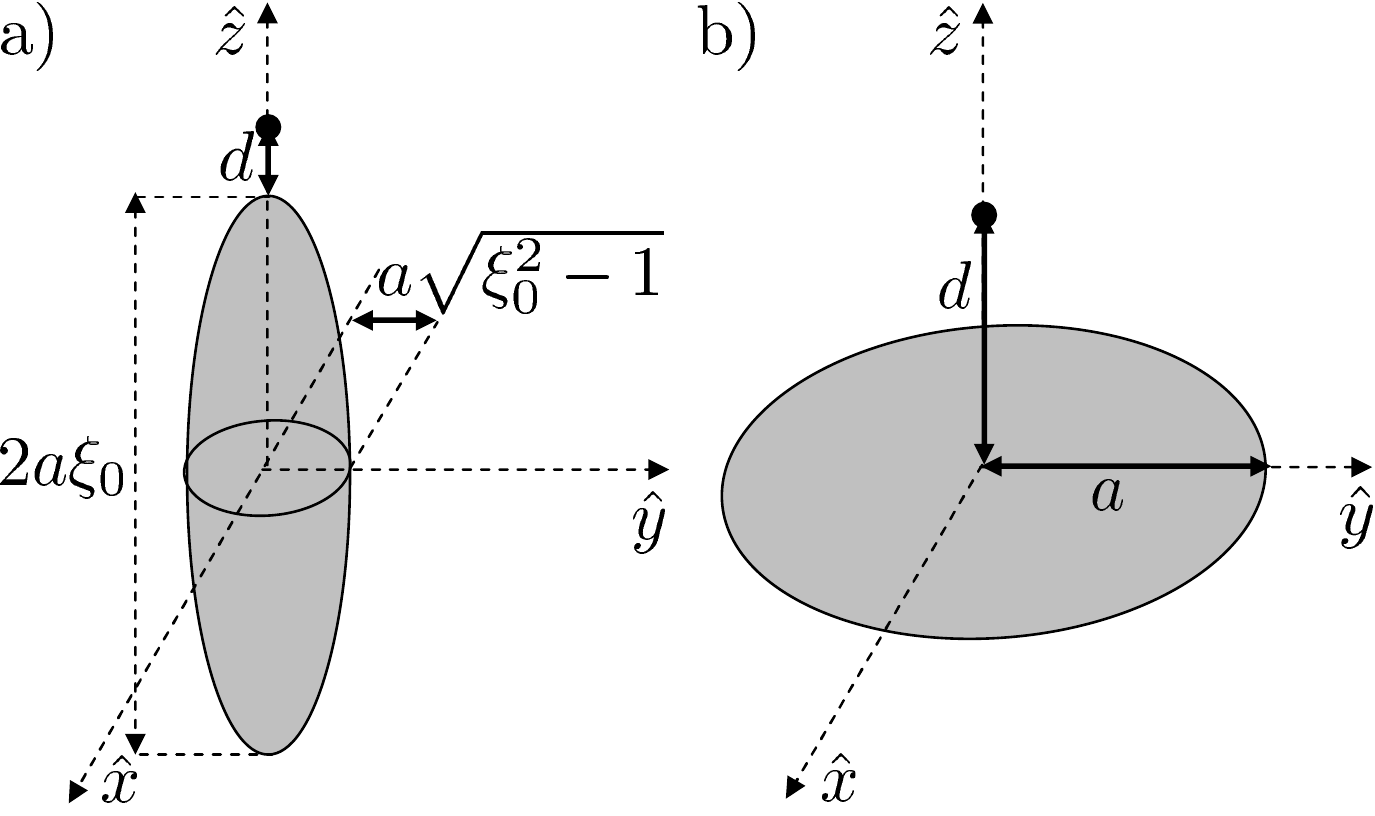}
\caption{(a) Needle-shaped electrode in prolate spheroidal coordinates for a surface $\xi > 1$. (b) Disc-shaped electrode in oblate spheroidal coordinates for a surface $\xi = 0$. The ion is represented by the black dot.
\label{fig:prolateoblate2}}
\end{figure}

The geometric factor, $\Lambda_{k}(\mathbf{r})$, is computed numerically for both needle and disc geometries by taking appropriate limits for prolate and oblate spheroids, respectively. We perform the sum over $l,l'$ by truncating at $l+l'=2l_{0}$. Similar to the case of the sphere, we define truncation at $l_0$ to correspond a patch size of $\theta_{\zeta}=2/l_{0}$. The lowest order non-vanishing term for the $\xi$ ($\eta$) mode occurs at $l_{0}=0$ ($l_{0}=1$). In both cases, scaling is evaluated with respect to distance $d$ along the $\hat{z}$ axis, as shown in Fig.~\ref{fig:prolateoblate2}.

\subsubsection{Prolate spheroid: needle}
\begin{figure}
\includegraphics[width=0.9\columnwidth]{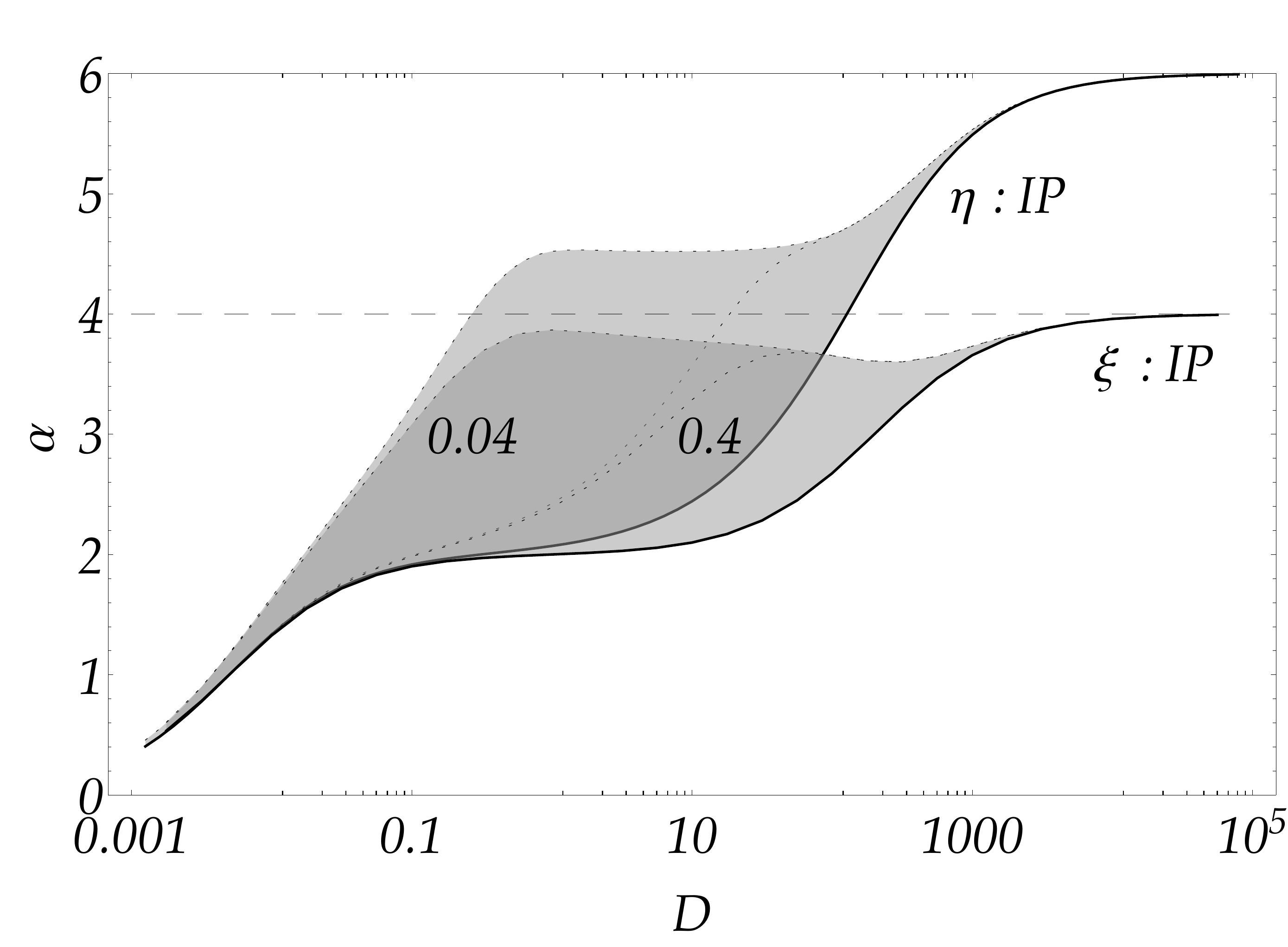}
\caption{Scaling exponent $\alpha$ as a function of dimensionless distance $D=d/r_\mathrm{prolate}$ above a needle of radius $r_\mathrm{prolate}=0.0100005\times a$ and half-length $100\times r_\mathrm{prolate}$ for both normal ($\xi$) and transverse ($\eta$) modes. The labeled solid lines are for the IP limits. The dotted lines correspond to intermediate patches sizes $\theta_{\zeta}=0.4,0.04$ evaluated via Eq.~\ref{eq:SEomegaProlate} by truncating the expression at $l+l'=2l_{0}=4/(\theta_{\zeta})$. The dashed line of $\alpha=4$ is plotted for reference.
\label{fig:GLOW_SAFEF_Needle}}
\end{figure}

The general features of the geometric factor for a prolate spheroid are illustrated through an example geometry with an aspect ratio, half-length/radius~$=100$. The high aspect ratio mimics that of needle-shaped electrodes as illustrated in Fig.~\ref{fig:needle_bispherical_sphere} (a) and (b). Specifically, we model the coordinate surface $\xi_{0} = 100/(3\sqrt{1111})\approx1.00005$. This represents a spheroid with a half-length of $\xi_0 a_\mathrm{prolate}\approx a 1.00005$ and a radius, $r_\mathrm{prolate}\approx a 0.0100005$, at the widest point. All dimensions are modeled in units of $a$ and we set $\mathbf{r}_{0}=\xi_{0}\hat{z}$, $\hat{d}=\hat{z}$, $\eta = 1$ in the geometric factor of Eq.~\ref{eq:geometricfactor}. With this choice of geometry $\hat{\xi} = \hat{z}$, and $d$ refers to distance above the top of the needle [c.f. Fig.~\ref{fig:prolateoblate2}~(a)]. The dependence of the scaling exponent, $\alpha$, with respect to the dimensionless distance, $D=d/(r_\mathrm{prolate})$, for the $\xi$ and $\eta$ modes are plotted in Fig.~\ref{fig:GLOW_SAFEF_Needle}.

The qualitative behavior of the scaling exponent for a prolate needle largely imitates that of the sphere: In the limit of $D\gg1$, the $\xi$ mode converges to $\alpha=4$ and the $\eta$ mode to $\alpha=6$, while in the limit of $D\ll1$, both modes exhibit convergence toward $\alpha=0$. Compared to the results for the sphere, however, additional features have arisen. The brief plateau in $\alpha$ that occurs for intermediate $0.1<D<100$ is a geometric effect due to the elongated shape of the needle. For lower aspect ratios we find that the region of this plateau narrows and that convergence toward $\alpha=0$ becomes evident already at larger values of $D$. We note that the bump in $\alpha$, for the $\xi$ mode around $D\sim1$, is also present in the exact analytic solution for spheres, and is not a numerical artifact. Such non-monotonic behavior is not entirely surprising as this region represents the cross-over between the two distinct regimes of infinite and infinitesimal ion-surface distance.

Similarly to the results of the sphere, we find, for the $\xi$ mode in the region explored by the Deslauriers et al. needle experiment ~\cite{Deslauriers2006}, that the possible values of $\alpha$ are bounded to an interval below $4$ but consistent with the $\alpha=3.5\pm0.1$ reported by their experiment.

\subsubsection{Oblate spheroid: disc}
The most extreme oblate spheroid is a disc with a radius of $\xi_{0} = a$ and zero thickness. This geometry may be used to model the surface electrode ion trap~\cite{Chiaverini2005a, Wesenberg2008, House2008}, currently receiving much attention due to its potential in quantum information science~\cite{Kielpinski2002} -- a field that experiences particular sensitivity to electric field noise~\cite{Epstein2007}. Considering this geometry furthermore extends the work of Dubessy et al.~\cite{Dubessy2009} to finite planes. 

For our numerical evaluation, we set $\mathbf{r}_{0}= 0$, $\hat{d}=\hat{z}$, and $\eta = 1$ in the geometric factor of Eq.~\ref{eq:geometricfactor}. With this choice of geometry, $d$ refers to distance above the disc origin [c.f. Fig.~\ref{fig:prolateoblate2}~(b)]. A complication occurs when evaluating $\alpha$ for the disc: As in the case of the hole trap, where the thin edge caused a logarithmic divergence in $\alpha$, a similar effect arises here. This is resolved by restricting the region of integration over $\eta$ in the $\beta^{\pm}_{ll'm}$ coefficient to $(-1,-\delta), (\delta,1)$, where $0<\delta \ll 1$ to avoid the edge. Here we have used $\delta = 0.1$.

\begin{figure}
\includegraphics[width=0.9\columnwidth]{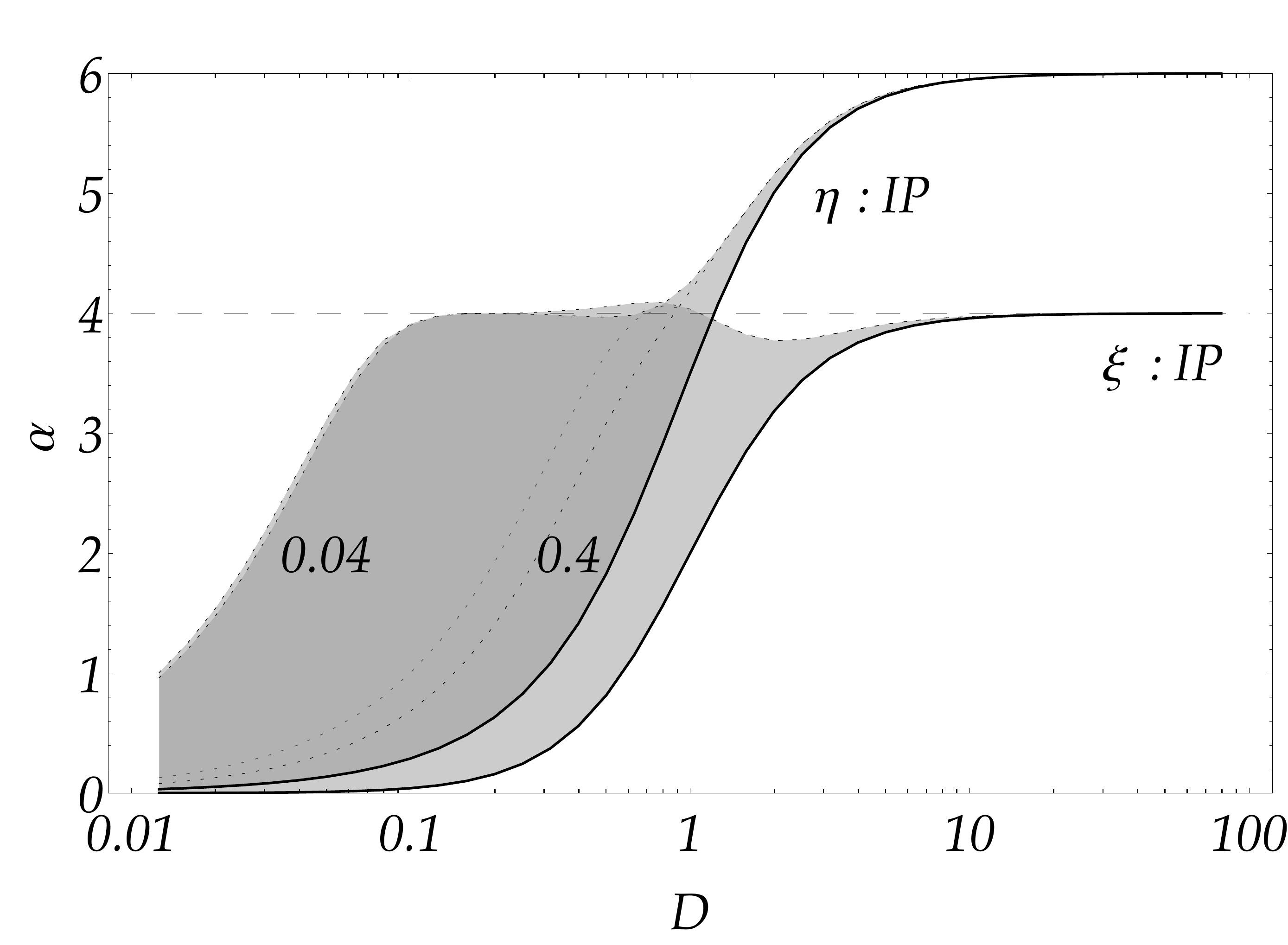}
\caption{Scaling exponent $\alpha$ as a function of dimensionless distance $D=d/a$ above a disc of radius $a$ for both normal ($\xi$) and transverse ($\eta$) modes. The labeled solid lines are for the IP limits. The dotted lines correspond to intermediate patches sizes $\theta_{\zeta}=0.4,0.04$ evaluated via Eq.~\ref{eq:SEomegaOblate} by truncating the expression at $l+l'=2l_{0}=4/(\theta_{\zeta})$. The dashed line of $\alpha=4$ is plotted for reference.
\label{fig:GLOW_SAFEF_Disc}}
\end{figure}

The scaling exponent, $\alpha$, for the $\xi$ and $\eta$ modes of the disc are plotted in Fig.~\ref{fig:GLOW_SAFEF_Disc}. The qualitative behavior is similar to the prolate needle geometry and, in some limits, to the infinite plane: In the limit of $D=d/a\ll1$, e.g., convergence toward $\alpha=0$ is observed, as expected in this limit where all surfaces appear infinite regardless of their exact geometry. In the limit of $D\gg1$ the $\xi$ mode converges to $\alpha=4$, while the $\eta$ mode converges toward $\alpha=6$, once again signifying a strong departure from the results of the infinite plane model.


\section{Conclusions}\label{sec:Conclusions}
We have presented an analytic model for electric field noise from fluctuating patch potentials starting from Laplace's equation and its Green's function solution. Beyond geometries for which the Green's function is readily obtained, our model uses an eigenfunction expansion and we employ this to analyze geometries of relevance to current ion trapping technology.

At distance scales that are relevant to typical ion traps, prior works have collectively established a scaling for electric field noise with surface distance, $d$, of $d^{-\alpha}$, with $\alpha=4$. While our model is in agreement with those models in a number of scenarios, for certain parameter regimes we observe a strong dependence on the finite geometry as well as on the patch size, leading to a departure from the $\alpha=4$ scaling. Moreover, we consider the effect of electric field noise on motional modes of a trapped ion that are both normal and transverse relative to the electrode surface studied. Significant differences between the two orthogonal modes are predicted for the more extreme geometries such as e.g. needle shaped electrodes. 

It would be of interest to confirm these predictions experimentally, and a suitable geometry could be either the needle trap of Deslauriers et al.~\cite{Deslauriers2006} or the recently developed stylus trap, which has also been proposed as a highly sensitive electric field probe~\cite{Maiwald2009}. Combined with techniques for varying the ion-surface distance, $d$, \emph{in-situ}~\cite{Herskind2009,Kim2010}, such systems could be used for detailed tests of our model. 

Although the geometries we have considered are mostly finite, not all are representative of real ion traps. In particular, the needle trap of Deslauriers et al.~\cite{Deslauriers2006} was approximated by a single needle rather than two as in the actual experiment. To what extent this influences the scaling is difficult to gauge intuitively, but it could potentially be investigated by modeling two spheres exactly in bispherical coordinates where Laplace's equation is separable.

More generally, the geometries considered in this work have been generic shapes for which solutions to Laplace's equation can be written out either through the appropriate Green's function or via an eigenfunction expansion. Real ion traps often differ from such geometries; however, if the electrostatic Green's function can be obtained numerically, the approach presented in this work could be extended to arbitrary geometries. In this respect we emphasize the method of the eigenfunction expansion. As it in principle allows for the study of arbitrary patch correlation functions, it may find use in modeling of  experimental scenarios relevant, for example, to Casimir force measurements and nanoscale surface probes, where the characteristic scale of the probe-surface distance becomes commensurate with the effective patch size.



\begin{acknowledgments}
This work was supported by the NSF Center for Ultracold Atoms and the IARPA SQIP program. P.F.H. is grateful for the support from the Carlsberg Foundation and the Lundbeck Foundation.
\end{acknowledgments}

%

\end{document}